# The Flux of Kilogram-sized Meteoroids from Lunar Impact Monitoring

R. M. Suggs[a*], D. E. Moser[b], W. J. Cooke[a], R. J. Suggs[a]
[a]NASA, Marshall Space Flight Center, Meteoroid Environment Office, Natural Environments Branch, EV44 Marshall Space Flight Center, Alabama 35812
[b]MITS/Dynetics, Marshall Space Flight Center, Meteoroid Environment Office, Natural Environments Branch, EV44 Marshall Space Flight Center, Alabama 35812
* Corresponding author. E-mail address: rob.suggs@nasa.gov (R.M. Suggs)

## Abstract

The flashes from meteoroid impacts on the Moon are useful in determining the flux of impactors with masses as low as a few tens of grams. A routine monitoring program at NASA's Marshall Space Flight Center has recorded over 300 impacts since 2006. A selection of 126 flashes recorded during periods of photometric skies was analyzed, creating the largest and most homogeneous dataset of lunar impact flashes to date. Standard CCD photometric techniques were applied to the video and the luminous energy, kinetic energy, and mass are estimated for each impactor. Shower associations were determined for most of the impactors and a range of luminous efficiencies was considered. The flux to a limiting energy of $2.5\times10^{-6}$ kT TNT or $1.05\times10^{7}$ J is $1.03\times10^{-7}$ km$^{-2}$ hr$^{-1}$ and the flux to a limiting mass of 30 g is $6.14\times10^{-10}$ m$^{-2}$ yr$^{-1}$ at the Moon. Comparisons made with measurements and models of the meteoroid population indicate that the flux of objects in this size range is slightly lower (but within the error bars) than flux at this size from the power law distribution determined for the near Earth object and fireball population by Brown et al. 2002. Size estimates for the crater detected by Lunar Reconnaissance Orbiter from a large impact observed on March 17, 2013 are also briefly discussed.

## 1. Introduction

The flux of kilogram-sized meteoroids is ill-determined due to their relatively low flux. Large collecting areas are needed to provide reasonable statistics for flux calculations. All-sky video systems used for fireball detection are limited to the roughly 10000 km$^2$ of atmosphere visible from their location and their sensitivity allows them to see down to sub-kilogram particles. Lunar impact monitoring utilizes the roughly $10^6$ km$^2$ collecting area (defined by the camera field of view) of the lunar surface to detect reasonable numbers of meteoroids in the 10s of grams to few kilograms size range. This is accomplished by observing the flash of light produced when a meteoroid impacts the lunar surface, converting a portion of its energy to visible light detectable from Earth.

The possibility of observations of meteoroid impacts on the Moon was discussed almost a century ago by Gordon (1921) and the implications of such observations for the existence of a lunar atmosphere were considered by La Paz (1938). As early as 1966, an attempt to observe lunar impacts during the Leonids yielded promising though unconfirmed results (Carpenter et al. 1967). In 1972, astronaut Harrison Schmitt observed a possible meteoroid impact from lunar orbit during Apollo 17 (NASA 1972) which may have been produced by a Geminid. The possibility of lunar impact flash detection from Earth was discussed and modeled by Melosh et al. (1993), Clark (1996), Beech & Nikolova (1998), Nemtchinov et al. (1998), and Shuvalov et al. (1999). Ortiz et al. (1999) made single telescope CCD observations of the Moon between 1997 and 1998 but could not conclusively distinguish between noise or seeing variations and a true impact flash. Unambiguous detections of lunar impacts began with video observations during the Leonid storm of 1999 (Bellot Rubio et al. 2000a; Dunham et al. 2000; Ortiz et al. 2000; Yanagisawa & Kisaichi 2002) and continued with the 2001 Leonids (Cudnik et al. 2002; Ortiz et al. 2002). The collected Leonid data constrained impact models and yielded insight on their thermal properties (Artem'eva et al. 2001). In addition to the Leonids, successful video observations of Geminid, Lyrid, Perseid, and Taurid impacts have been reported (Yanagisawa et al. 2006; Cooke et al. 2006, 2007; Yanagisawa et al. 2008; Moser et al. 2011). Observations made outside shower peak periods have also yielded impact flashes detailed in Ortiz et al. (2006), Cooke et al. (2007), and Suggs et al. (2008, 2011).

NASA's Marshall Space Flight Center (MSFC) implemented a video monitoring program to routinely observe the Moon for impact flashes using multiple telescopes in early 2006. This has resulted in the observation of over 300 lunar impacts in roughly 7 years. This paper summarizes the results of the first 5 years of lunar impact monitoring at MSFC and updates previous results presented in Suggs et al. (2008, 2011). Consistent observational practices and careful photometric calibration yield a dataset of 126 impact flashes, the largest and most homogeneous to date. The monitoring technique, photometric calibration, and selection of the best data from the program are described. Calculation of impact kinetic energy, association with meteor showers, and calculation of impactor mass are discussed. The flux to a limiting energy and a limiting mass are compared to measurements and a model for other size ranges.

## 2. Observations

### *2.1 Method*

The earthshine hemisphere of the Moon is observed between 0.1 (crescent) and 0.5 (first quarter) phase and 0.5 (last quarter) to 0.1. The video field of view is oriented with the equator along the vertical axis and limb in the field of view. This maximizes the lunar surface area observed and minimizes glare from the sunlit hemisphere. Evening observations (waxing phase) cover the western or leading hemisphere while morning observations (waning phase) cover the eastern or trailing hemisphere. Figure 1 shows a Lyrid impact on 22 April 2007 at 03:12:21 UT (impact #26 in Table 1) and illustrates the video field of view. Lunar surface features are easily visible in earthshine and are used to determine the location of the flash.

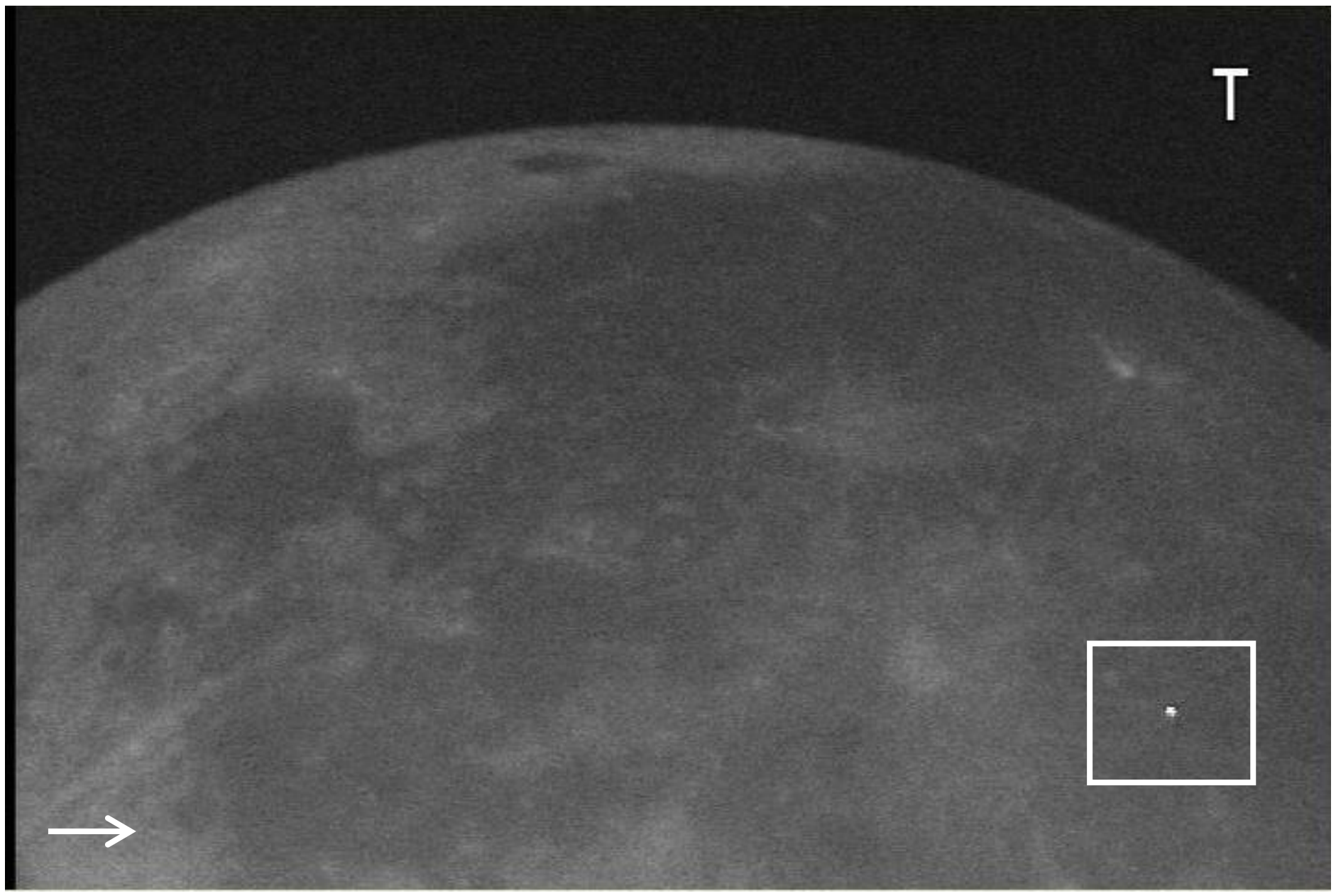

**Fig. 1.** Lyrid impact flash #26 on 22 April 2007 at 03:12:21 UT. The arrow indicates the direction of selenographic north. The horizontal field of view extent is approximately 20 arcminutes.

Schmidt-Cassegrain telescopes 0.35 m (14 inch) in diameter were used to observe the Moon although some observations were made with a 0.5 m Ritchey Chretien instrument (from 23 January 2008 to 23 January 2010). Focal reducers were used to provide a field of view on the video chip of approximately 20 arcminutes along the long axis of the frame. This field of view covers approximately $3.8\times10^6$ $km^2$ of the lunar surface. The cameras were "1/2 inch" format NTSC video based on the Sony EXview HAD CCD™ chip. Cameras based on this CCD were chosen because of the high sensitivity of the Hole Accumulation Diode (HAD) and EXview microlens technology. This camera/telescope combination gave a limiting stellar R magnitude of approximately 10.5. The frame rate was standard 30 per second with interleaved 1/60 s fields. The video signals were digitized and recorded on PC harddrives for subsequent flash searches and photometric analysis. The digitizers performed mild data compression which did not significantly affect photometry as is evidenced by the near zero average error and 0.2 magnitude standard deviation determined for the ensemble of comparison stars. All photometry was based on the 1/60 s fields by extracting even and odd rows from the video frames.

The cameras were set to manual gain with electronic shutter control off. The cameras and settings used are described in Appendix A. All reported flashes were confirmed using at least 2 telescopes to discriminate against cosmic ray flashes in the CCDs. Two telescopes were operated at MSFC's Automated Lunar and Meteor Observatory (Minor Planet Center designation H58: 34.66° N, 86.66° W). A third 0.35 m telescope was operated near Chickamauga, Georgia (34.85° N, 85.31° W), 125.5 km from MSFC, beginning in September 2007 to eliminate orbital debris sunglints up to geosynchronous altitude. Correlated observations from this telescope showed conclusively that the flashes observed at MSFC could not be from orbital debris. Even without this third telescope, any satellite or debris sunglint lasting more than a few frames shows motion across the field of view unlike the stationary lunar impact flashes.

For some observations in 2009 and 2010 an InGaAs near-infrared (0.9 – 1.7 micron) video camera was used on one of the telescopes. It proved useful for confirmation but not for photometry due to persistence issues.

### *2.2 Flash Detection and Aperture Photometry*

Impact flash detection was performed using LunarScan (Gural 2007). The software looks for pixels that exceed the standard deviation over the mean image by a factor of 3.5. The mean and standard deviation are tracked on a frame by frame basis using a first order response filter for each pixel. A spatial correlation filter looks for 3 rows of exceedances which approximates the optical system point spread function. Candidate impact flashes are manually correlated with video recorded from the second and, when available, third telescope to reject any cosmic rays or satellite glints. Figure 2 shows a false color sequence of video frames of the impact shown in Figure 1. The LunaCon program (Swift et al. 2008) was used to extract the aperture photometry data for the flashes and for stars near the limb passing through the field of view (the field stars) as well as to display the lunar contrast which was used to exclude periods of clouds and poor photometric quality. This information was used to select the flashes and observation time spans as described in Section 4.1. LunaCon was also used to determine the lunar area within the field of view.

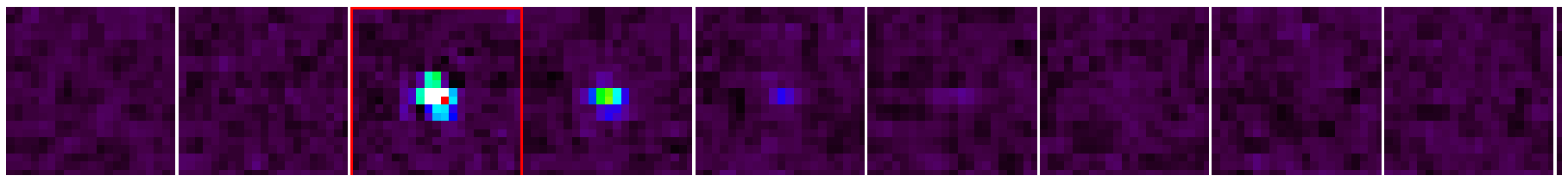

**Fig. 2.** Video frame sequence for flash #26 on 22 April 2007 at 03:12:21 UT. Each square in the time series represents 1/30 s and is approx. 35”×35”.

## 3. Photometric Calibration

Standard aperture photometry was applied to the flashes and the field and reference stars used for calibration. Images were flatfielded using skyflats. The impact flash video field (1/60 s) with the largest signal was used in these analyses. Ernst and Schultz (2005) showed that the peak luminous energy in their hypervelocity gun tests occurred on the 10 – 20 μs time scale. The video exposure time is 1000 times longer even after adjustment for the impactor diameter to velocity ratio they used. Bouley et al. 2002 suggest that luminous efficiencies determined using the entire light curve for Leonids might indicate differences in the impactor or the lunar soil at the impact site. Yanagisawa and Kisaichi 2002 proposed that the prolonged afterglows they observed for impact flashes were due to thermal emission from droplets of lunar soil vaporized and recondensed in flight. Thus the best estimate of impact kinetic energy comes from the shortest exposure time rather than the entire light curve. Subsequent video frames give good information on the rate of cooling of ejecta material (Bouley et al. 2012) but no useful information on the kinetic energy of the impact.

Two types of stellar calibration sources were available: field stars, those that passed by the Moon in the video field of view, and reference stars, a set of 31 stars observed over 2 nights in November 2012. The former were used to determine the photometric zero point each night and to determine the extinction coefficient if there were 3 or more stars at various airmasses available. The latter were used to determine the color correction term for the stars, to determine an extinction coefficient and to investigate the error due to scintillation effects. Details of the photometric calibration and error determination are given in Appendix A.

## 4. Analysis and Results

### *4.1 Flash and Observation Time Span Selection*

For this analysis the first 5 years of observations ending in August 2011 were considered. Since passing clouds during the recording period can drastically affect the limiting magnitude and hence the flux calculations, only periods of good transparency were used in the observing time calculations. The contrast between the lunar surface (earthshine) and the sky are reported by LunaCon and are used to judge the photometric quality of the data and select the "photometric" periods. Any flashes observed during periods of poor transparency were discarded. This criterion yielded the 126 flashes in 147 observing sessions, a total of 266.88 hr, used in this study. Figure 3 shows the location of these impact flashes.

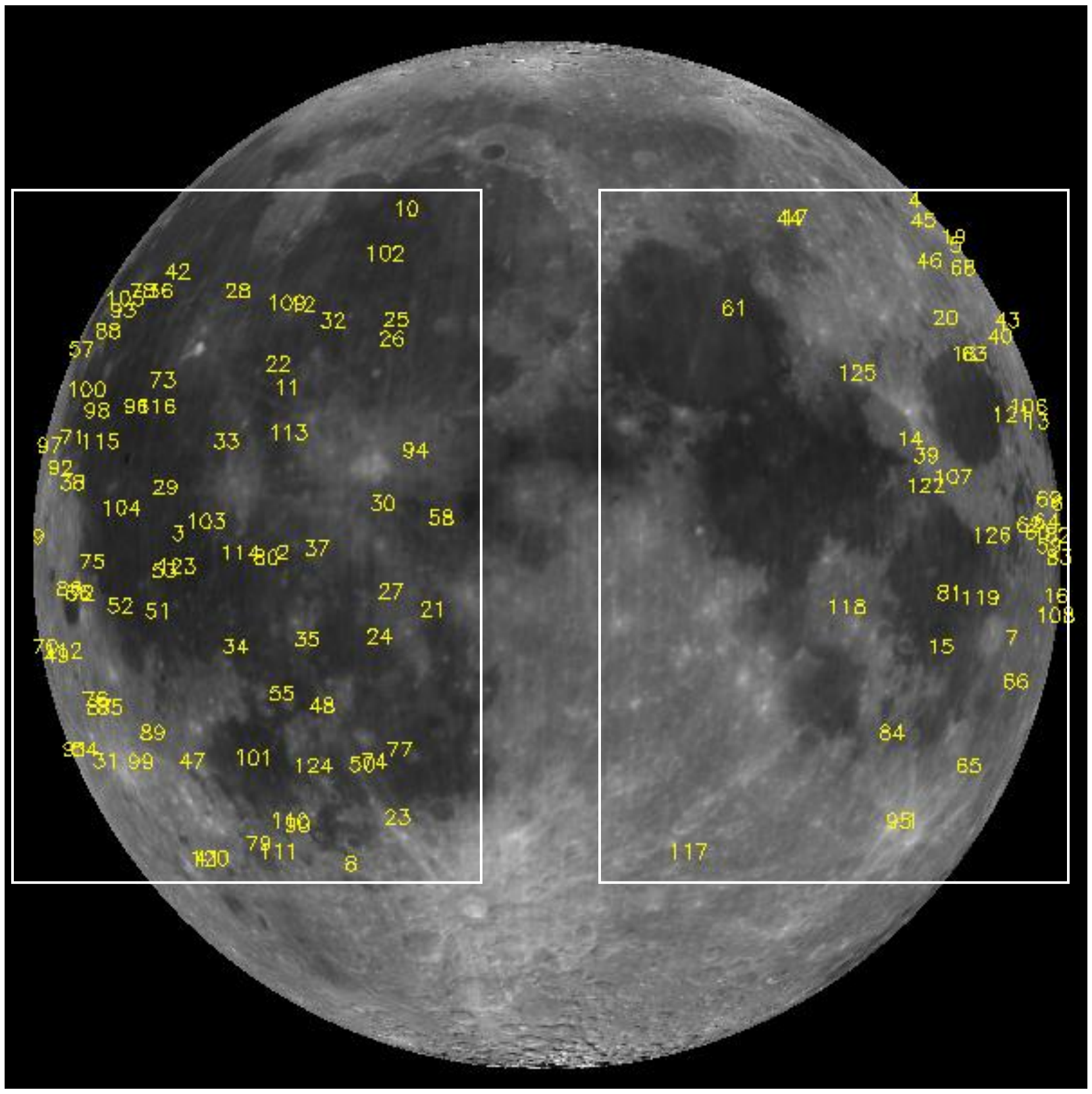

**Fig. 3.** Locations of the 126 impact flashes in this study. The approximate extent of the camera field of view during the waxing and waning phase observations is shown.

Table 1 shows the flash dataset, listing the date, time, and solar longitude of observation, the number of telescopes confirming the detection, the flash location, the air mass at the time of observation, the flash peak R magnitude with photometric uncertainties (see Section A.2), and peak luminous energy (discussed in Section 4.3). Flash magnitudes range from 10.42 to 5.07, with corresponding luminous energies between $3.68\times10^3$ J and $5.08\times10^5$ J.

INSERT TABLE 1 HERE

*4.2 Shower Identification*

Figure 4 shows the impact flash rate in 2 degree bins of solar longitude calculated from the number of impact flashes and number of observation hours in each bin. This simple calculation does not consider the visibility of the radiant from the flash impact point but the correlation of the peak rates with meteor showers is evident.

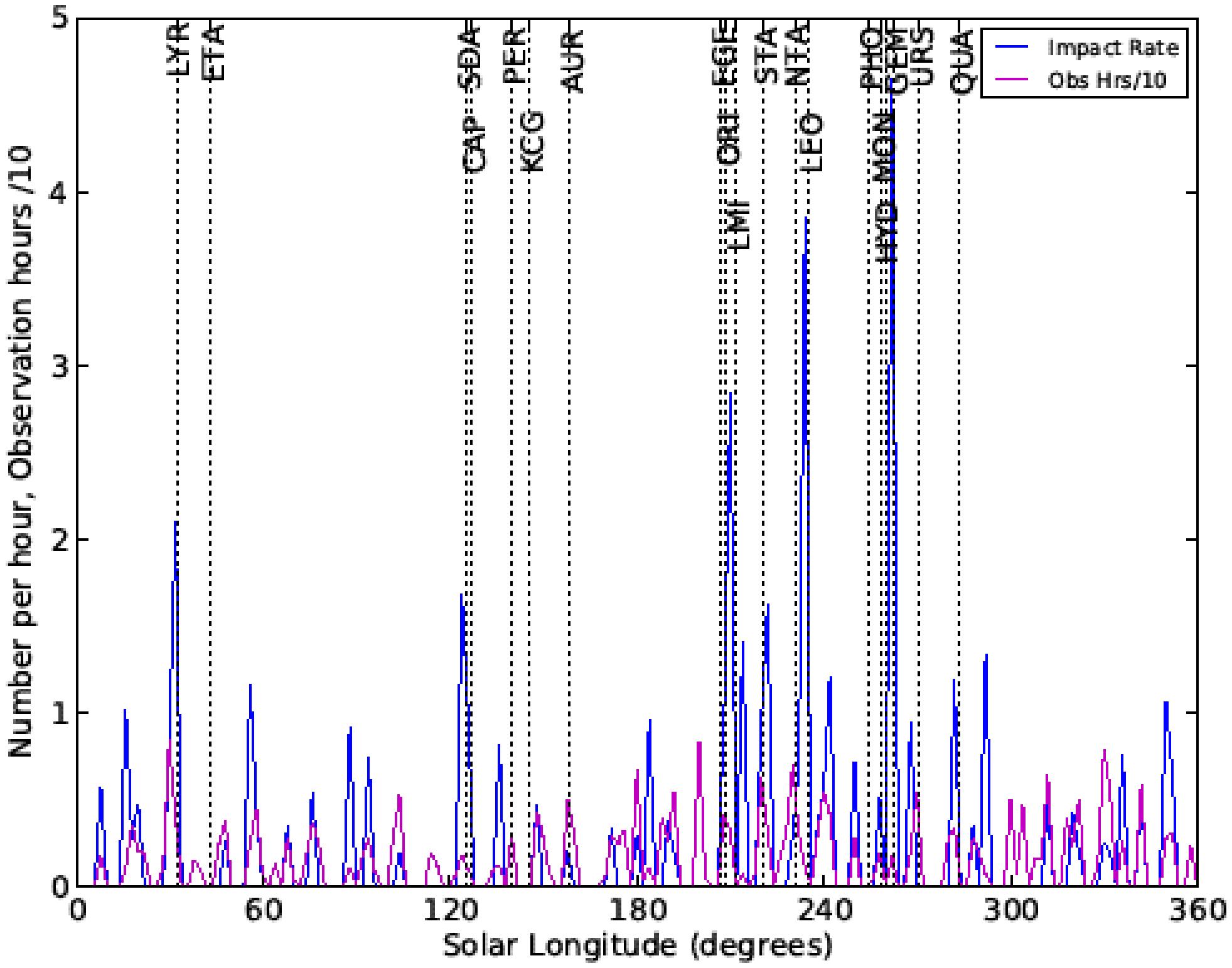

**Fig. 4.** Impact flash rate (number of flashes per observation hour) in 2 degree bins in solar longitude (blue curve). Shower peak times are indicated by dashed vertical lines. The total number of observation hours divided by 10 is shown by the magenta curve.

Note the strong shower correlation with the largest peaks in activity. The absence of any Perseids is likely due to the small number of hours of observation, less than 2 hours, during the peak of the shower. This is due to a combination of Alabama summertime weather and poor visibility of the radiant from the portion of the Moon imaged on the few clear August nights. The asymmetry in impact rates evident in Figure 3 is real and is due to the visibility of different meteor shower radiants and sporadic sources from the area in the field of view. A detailed analysis of this asymmetry is the subject of future work.

It is impossible to uniquely assign any given impact flash to a particular meteor shower or sporadic source. The likelihood that a particular impact flash is due to a shower meteoroid can, however, be related to three parameters. The ZHR (Zenithal Hourly Rate) is a measure of the relative flux of meteoroids in a shower. Showers with higher ZHRs are more likely to produce more impact flashes. The time of the impact compared with the time of the peak and the length of the shower is also important. These are usually expressed in terms of the longitude of the sun, $\lambda$, at that time. The third parameter is the distance of the impact point from the sub-radiant point on the Moon. There are more impacts per square km at points directly beneath the radiant than at points near the radiant visibility limit. To combine these parameters a Figure of Merit (*FOM*) was calculated for each impact flash according to

$$FOM = ZHR \times FOM_{time} \times FOM_{geom} \tag{1}$$

where

$$FOM_{time} = 1 - (\lambda_{max} - \lambda_{flash}) / (\lambda_{max} - \lambda_{beg}) \tag{2}$$

for flashes before the shower maximum and

$$FOM_{time} = 1 - (\lambda_{flash} - \lambda_{max}) / (\lambda_{end} - \lambda_{max}) \tag{3}$$

for flashes after maximum. $\lambda_{flash}$ is the solar longitude at the time of the impact flash, $\lambda_{max}$, $\lambda_{beg}$, and $\lambda_{end}$, and are the solar longitudes for the time of shower maximum, beginning, and end. Asymmetric showers with different rise and decay times are accommodated with this technique.

The geometric Figure of Merit is calculated from

$$FOM_{geom} = \cos\,\theta \tag{4}$$

where $\theta$ is the angle between the sub-radiant point and location of each impact flash. This angle was determined using calculations from JPL Horizons (http://ssd.jpl.nasa.gov/?horizons) and Analytical Graphics Systems Tool Kit (http://www.agi.com/). These calculations were also used to compute the impact speed incorporating the Moon's motion relative to the shower meteoroids.

Several meteor shower databases were considered for use in determining the most likely association of a given impact flash with a shower. Modern shower catalogs like Brown et al. (2010) and Jenniskens (2006) are based on radar and/or video measurements of meteors in the Earth's atmosphere. These techniques are sensitive to meteoroids which are much smaller than those observed impacting the Moon and the activity profiles for small meteoroids can be different from those for larger. A list of showers based on the larger meteoroids detectable by less sensitive photographic and visual observations is available in Cook (1973) which is based primarily on the Harvard Super Schmidt photographs. By using this list extrapolation to the larger meteoroids which cause observable impact flashes is over a smaller range of sizes so assumptions about the mass index (size distribution of meteoroids) are less important. Since mass indices in this size range are not available for all of the showers considered no attempt was made to correct the ZHR for size distribution. A shower catalog was constructed for this study from Cook's working list. ZHRs were added from Jenniskens (1994) because Cook's catalog had many missing ZHRs. This catalog is given in Table 2.

INSERT TABLE 2 HERE

The *FOM* technique yielded meteor shower associations for 79 of the 126 flashes. The dataset contains 10 Lyrids, 2 eta Aquariids, 7 South delta Aquariids, 13 Orionids, 15 South Taurids, 12 North Taurids, 4 Leonids, 2 December Monocerotids, 9 Geminids, 1 Ursid, and 4 Quadrantids. The remaining 47 flashes are unclassified, either because the *FOM* did not convincingly indicate one shower over another (i.e. 2 showers produced similar *FOM*s) or no shower was indentified. Of those with no shower identified, these may be associated with minor showers not included in our database, unknown showers, or sporadics. Meteor shower associations are shown in Table 3.

### *4.3 Impact Energy Calculation*

While it is the impact kinetic energy that needs to be determined, it is the luminous energy that is observed. The luminous energy of each flash can be calculated from the peak R magnitude. Bessell et al. (1998) used stellar atmosphere models to generate bolometric corrections for the Johnson-Cousins filter system including the R filter. The bolometric (total luminous) energy is given by

$$E_{lum} = f_\lambda \, \Delta\lambda \; f \; \pi \; d^2 \; t \quad \text{Joules} \tag{5}$$

where

$$f_\lambda = 10^{-7} \text{ x } 10^{-(R + 21.1 + zp_R)/2.5} \quad \text{J cm}^{-2}\text{ s}^{-1}\text{ Å}^{-1} \tag{6}$$

$\Delta\lambda$ is the width of the filter passband (1607 Å), $d$ is the distance to the Moon ($3.844\times10^{10}$ cm), and $t$ is the exposure time (0.01667 s). The zero point magnitude for the R filter, $zp_R$, is given as 0.555 in Bessel et al. (1998). Note that table A2 of that paper has the $zp(f_\lambda)$ and $zp(f_\nu)$ row labels switched. The factor $f$ is related to the solid angle of emissions from the observed impact plume. For a flash in free space $f$ is 4 and for a flash close to the lunar surface where all the radiation is emitted into $2\pi$ steradians it should be 2. Since the initial frame of the flash is analyzed here the emissions should be from near the surface so $f = 2$ is used.

To get the impact kinetic energy (*KE*) it is necessary to know the luminous efficiency which is the fraction of the impact energy emitted as light in the camera passband. This fraction is denoted $\eta$.

$$KE = E_{lum} / \eta \tag{7}$$

The luminous efficiency of the Leonids was determined by Bellot Rubio et al. (2000b). Swift et al. (2011) derived an expression for luminous efficiency as a function of speed, $v$, using hypervelocity impact tests and luminous efficiencies for the CCD video camera passband determined by Moser et al. (2011) for three meteor showers. This expression is

$$\eta = 1.5\times10^{-3} \exp\left(-(9.3\text{km/s})^2/v^2\right) \quad (8)$$

Since there is considerable uncertainty in the determination of luminous efficiency, the calculations were also performed with extreme values of $\eta = 5\times10^{-4}$ and $\eta = 5\times10^{-3}$ as used by Bouley et al. (2012). Figure 5 shows the kinetic energy for each flash including the photometric uncertainties (red error bars) described in Section A.2. The order of magnitude uncertainty in luminous efficiency is shown as blue error bars. Values for each flash are given in Table 3. Kinetic energies range from $7.36\times10^5$ to $1.02\times10^9$ J over the whole range of luminous efficiencies considered.

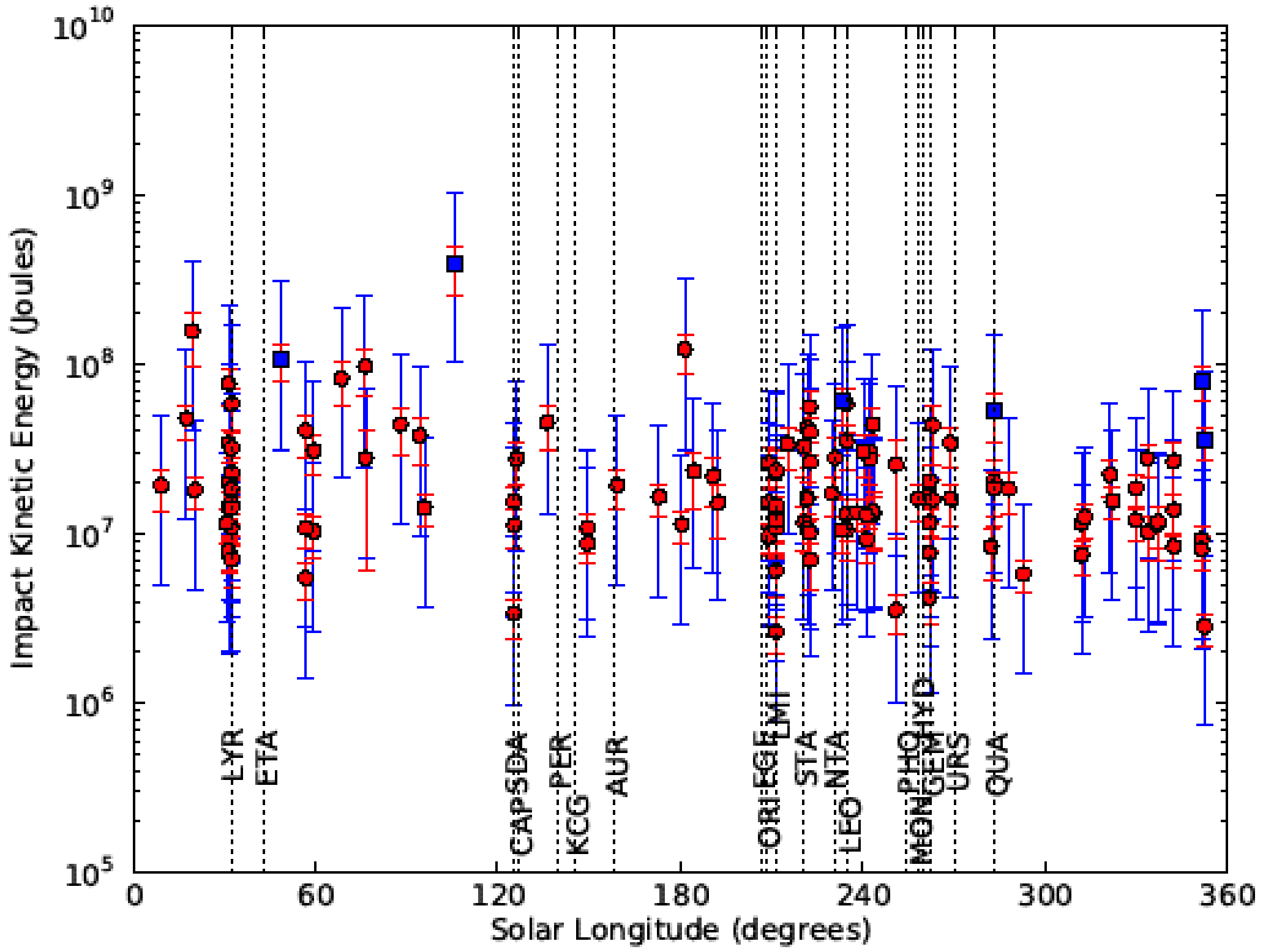

**Fig. 5.** Kinetic energy of each impact flash at its solar longitude. The red error bars indicate the uncertainty in the photometry and the blue error bars indicate the extremes of luminous efficiencies used in this study. The blue squares are flashes whose signals exceeded the dynamic range of the camera. Their values indicate a lower limit on the energy as described in the text in Appendix A.

### *4.4 Impactor Mass Calculation*

The mass of the impactor is easily calculated from the kinetic energy and speed using

$$M = 2\ KE\ /\ v^2 \tag{9}$$

Figure 6 shows the calculated masses with the photometric uncertainty and luminous efficiency extremes. Considering these extremes, masses range between 0.4 g and 3.5 kg. The values for each flash are given in Table 3. Shower impact speeds corrected for the Moon's velocity relative to the meteoroid velocity were used for those with shower associations. A speed of 24 km/s (McNamara et al. 2004) was used for the others. Note that no adjustment for gravitational focusing has been made because the effect is negligible for the Moon at typical meteoroid speeds (Oberst et al. 2012).

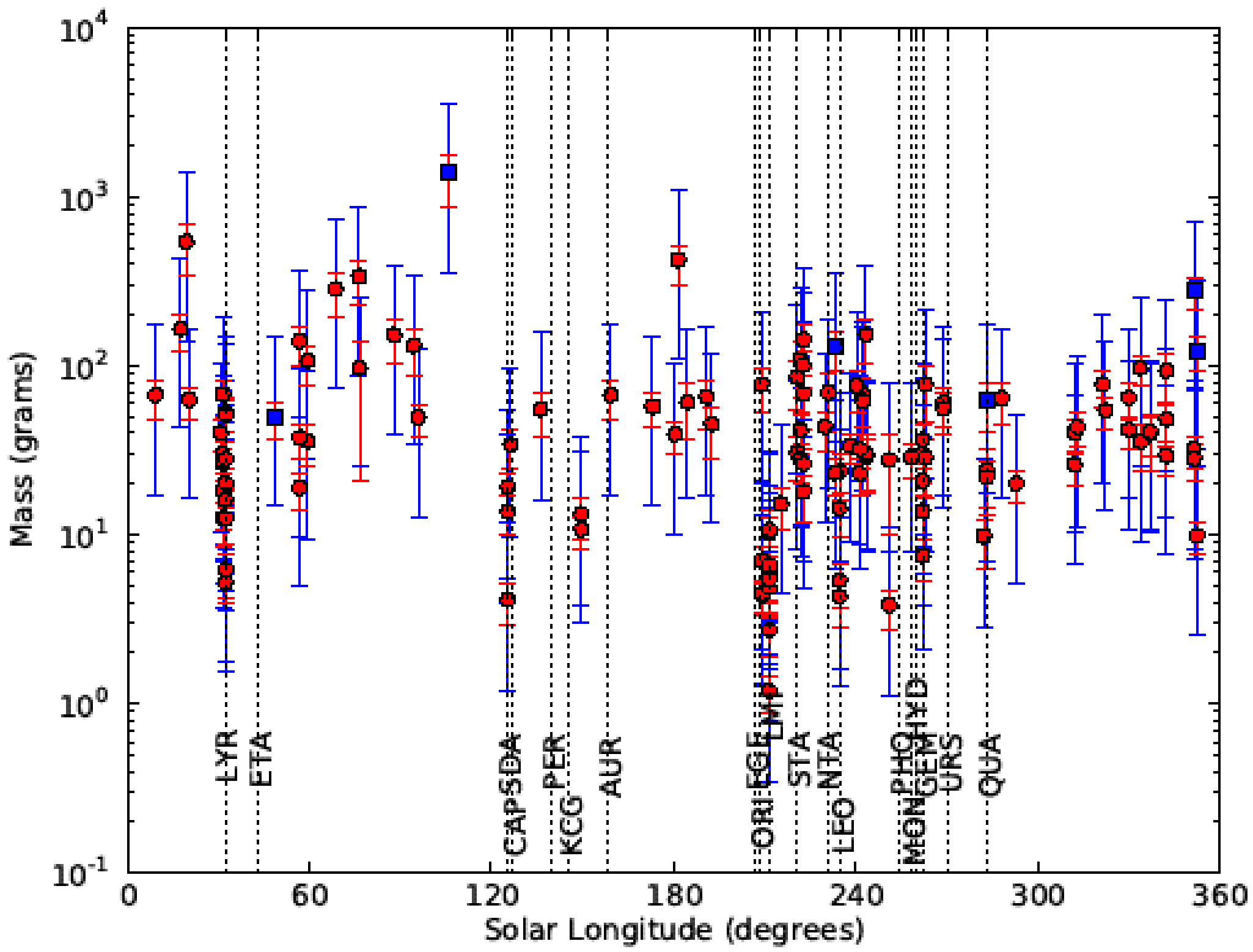

**Fig. 6.** The mass of each impactor with red error bars indicating the photometric uncertainty and blue error bars indicating the extremes of luminous efficiency used in this study. The blue squares are flashes whose signals exceeded the dynamic range of the camera. Their values indicate a lower limit on the mass as described in the text in Appendix A.

INSERT TABLE 3 HERE

*4.5 Flux Determination*

The flux of meteoroids can be expressed relative to a limiting energy or a limiting mass. Since the measured quantity is a stellar magnitude which is directly related to luminous energy, the limiting magnitude of the observations determines the limiting energy. The R magnitude distribution is given in Figure 7. The cumulative distribution in Figure 8 shows a turnover at R magnitude 9.0. Since the number of meteoroids increases with decreasing size by a power law, this turnover indicates the completeness limit for the observations.

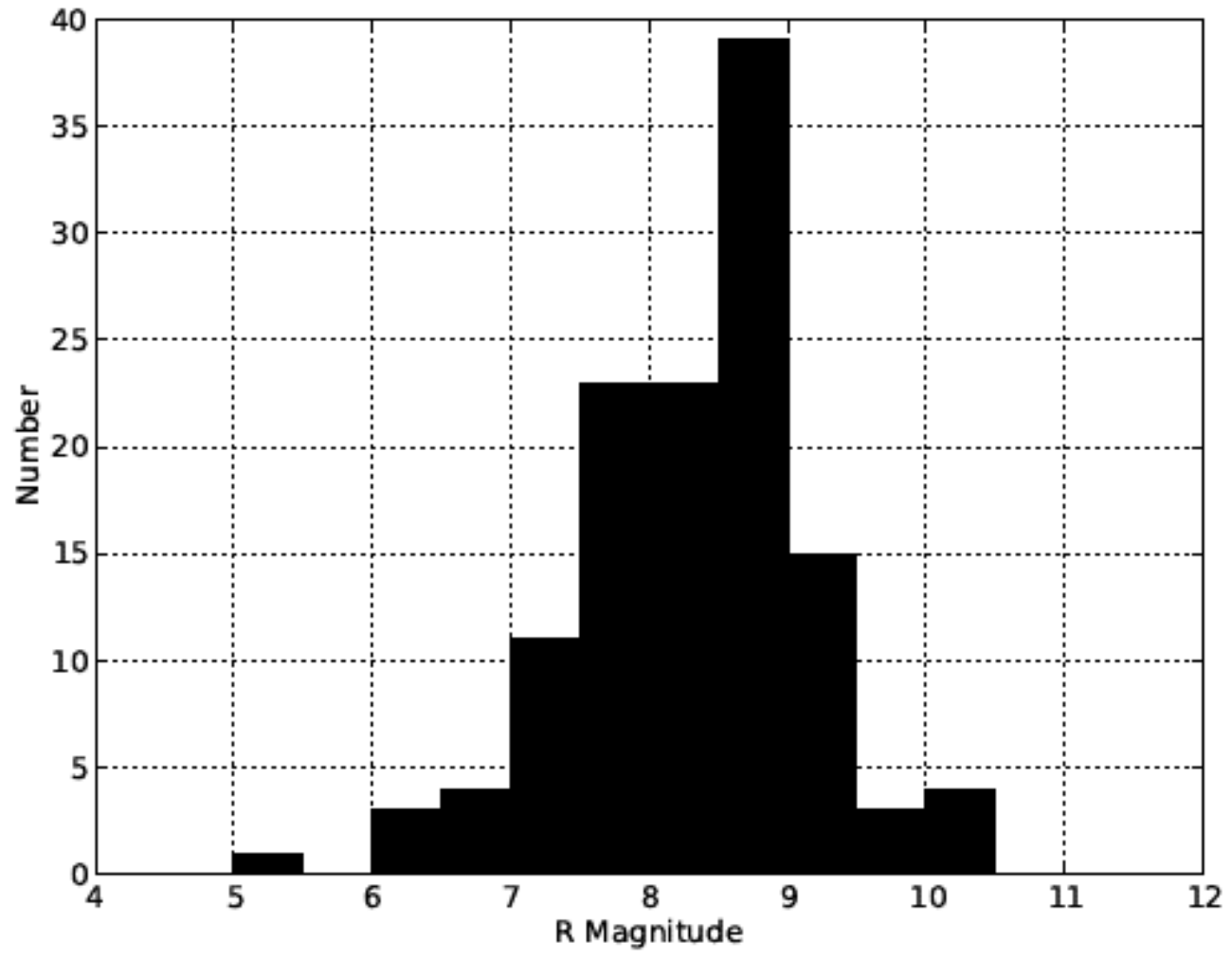

**Fig. 7.** Histogram of peak magnitudes of the impact flashes

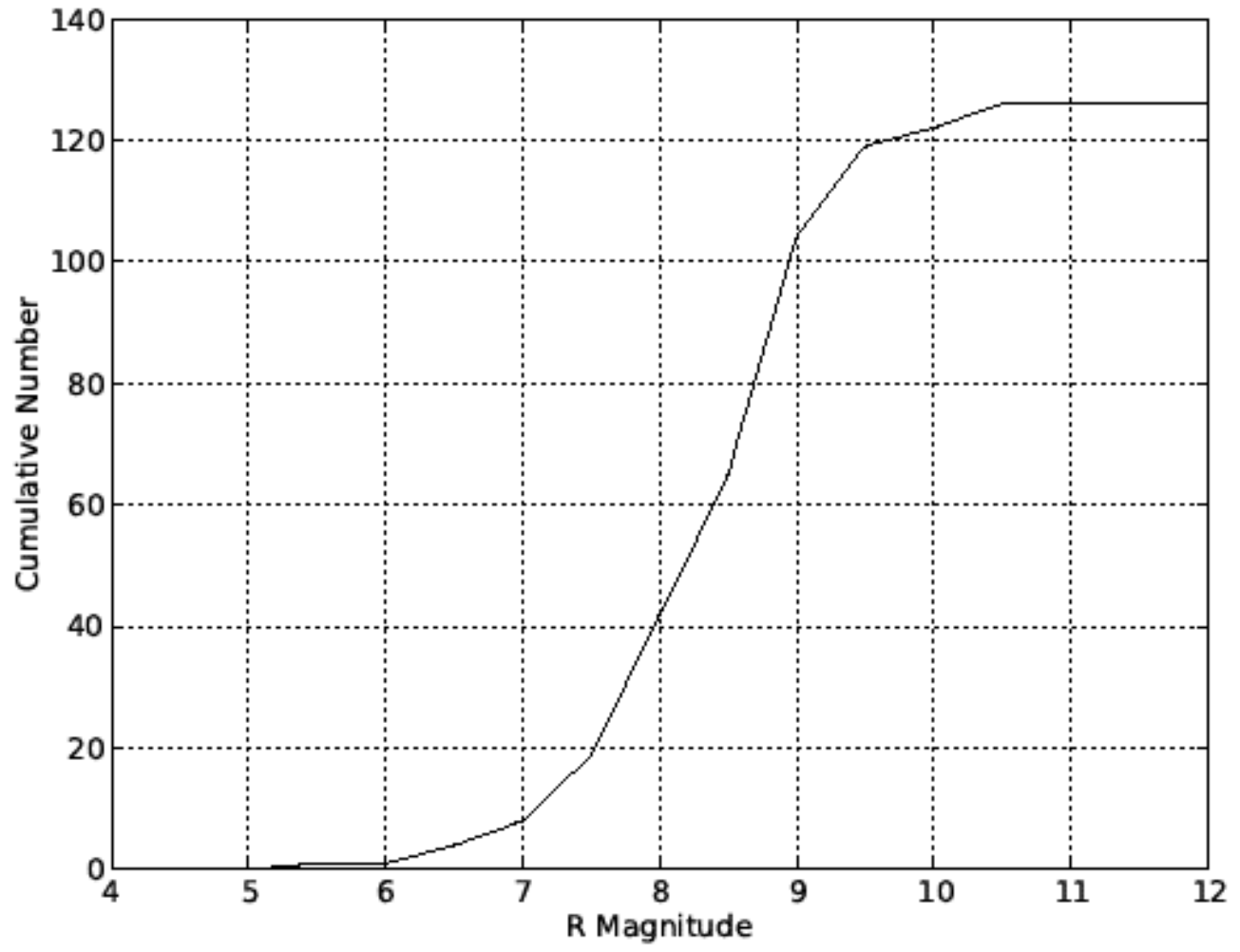

**Fig. 8.** Cumulative histogram of flash magnitudes indicating a turnover at magnitude 9 which is taken to be the completeness level of the data used in this study.

Examining the 126 flashes in the dataset, there are 104 flashes with magnitude brighter than or equal to 9.0. The flux is given by

$$F\,(m_R \leq 9.0) = N_{9.0}\,/\,(\tau A) \tag{10}$$

where $N_{9.0}$ = 104 is the number of impact flashes with magnitude $m_R$ less than or equal to 9.0, $\tau$ = 266.88 hr is the total duration of observations, and $A = 3.8\times10^6$ km$^2$ is the lunar collecting area. This calculation yields a flux of $1.03\times10^{-7}$ meteoroids per hour per km$^2$ on the Moon to a limiting magnitude of 9.0.

To compare this with Brown et al. (2002), this flux is converted to impacts on Earth's atmosphere per year to a limiting energy in terms of TNT equivalent (1 metric ton TNT = $4.18\times10^9$ J). The limiting magnitude of 9.0 corresponds to a kinetic energy of $1.3\times10^7$ J (assuming luminous efficiency of $1.29\times10^{-3}$ and energy scaling from equation 12 for a 24 km/s impact) which is $3.0\times10^{-6}$ kilotons of TNT at the Earth. Figure 9, after Brown et al. (2002), shows the cumulative flux and error bars due to uncertainty in luminous efficiency and number of flashes with energies at or above each flash after correction for the effect of gravitational focusing by the Earth. Gravitational focusing has two effects: it increases the impact speed and hence energy and it increases the impact parameter for capture which is essentially the collecting area of the Earth. These are both functions of the meteoroid speed. The energy correction was applied to each meteoroid energy and the error bars which represent the luminous efficiency uncertainty. The collecting area correction was computed for each meteoroid and the average was used in the flux calculation. The area factor, $H_F$, (from Jones and Poole 2007) is given by

$$H_F = A_G/A_{NG} = 1 + v_{\mathrm{esc}}^{2}\,/\,v^2 \tag{11}$$

Where $A_{NG}$ is the physical cross sectional area of the Earth, $A_G$ is the effective cross sectional area of the Earth taking gravitational focusing into account, $v_{esc}$ is the escape velocity of Earth (11.09 km/s at 100km altitude) and $v$ is the speed of the meteoroid relative to Earth prior to being accelerated. The average area correction factor was 1.14.

The kinetic energy correction factor is computed from the speed of a meteoroid at the top of the atmosphere due to gravitational acceleration. It is given by McKinley (1961) as

$$v_a^{2} = v^2 + v_{esc}^{2} \tag{12}$$

Where $v_a$ is the apparent velocity after gravitational acceleration.

This result agrees with the trend over many orders of magnitude in energy but falls slightly below the line at higher energies. As discussed in Appendix A, six of the most energetic impacts are saturated but those do not explain the downward trend with energy. There may be an unmodeled mass dependence in luminous efficiency. Bellot Rubio et al. (2000) state that the luminous efficiency varies little with mass but say that "there are some hints that this dependence might be significant". More work is needed in this area. The fireball data (red curve) indicates a turnover at these lower energies but seems to show a greater decrease than indicated by this lunar impact-derived curve, however, the error bars due to luminous efficiency uncertainty encompass the power law and fireball trend extrapolation. The fluxes determined by Ortiz et al. 2006 from 3 video impact flashes are also shown.

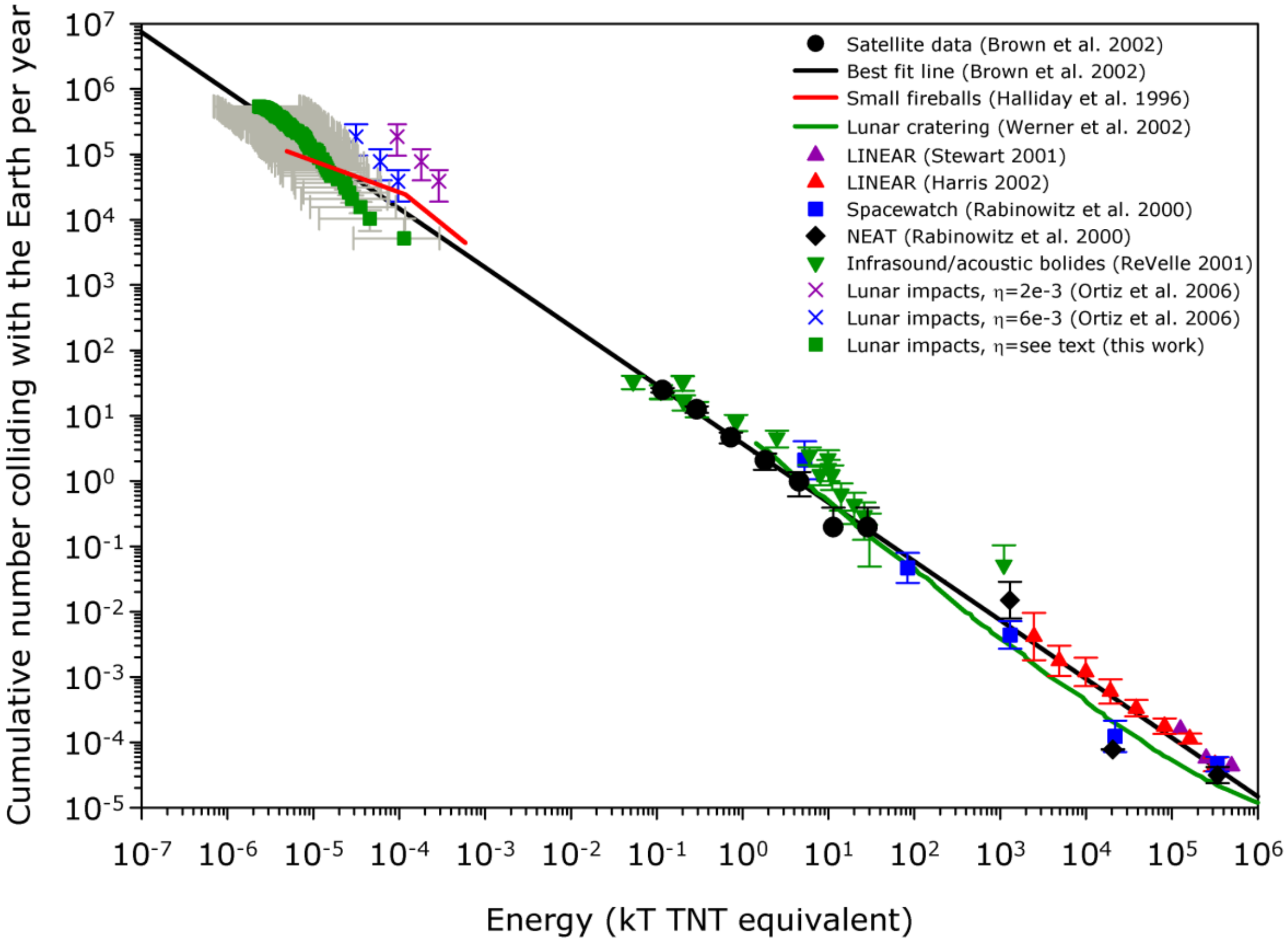

**Fig. 9.** Number of meteoroids versus energy striking Earth each year, after Brown et al. (2002). The cumulative flux for all energies greater than each impact flash are shown as the green squares with error bars indicating the uncertainty in luminous efficiency (horizontal bars) and the square root of the number of flashes (vertical bars). The lunar impact flash energies and Earth collecting area have been corrected for gravitational focusing.

Determining the limiting mass is not as straightforward because the detection limit is based on energy and the computed mass depends on impact velocity. Figure 10 shows the mass distribution in Table 3 which assumes the velocity from any shower associations or 24 km/s otherwise. Figure 11 is the cumulative distribution on a log scale. The turnover is around log mass of 1.5 corresponding to 30 g.

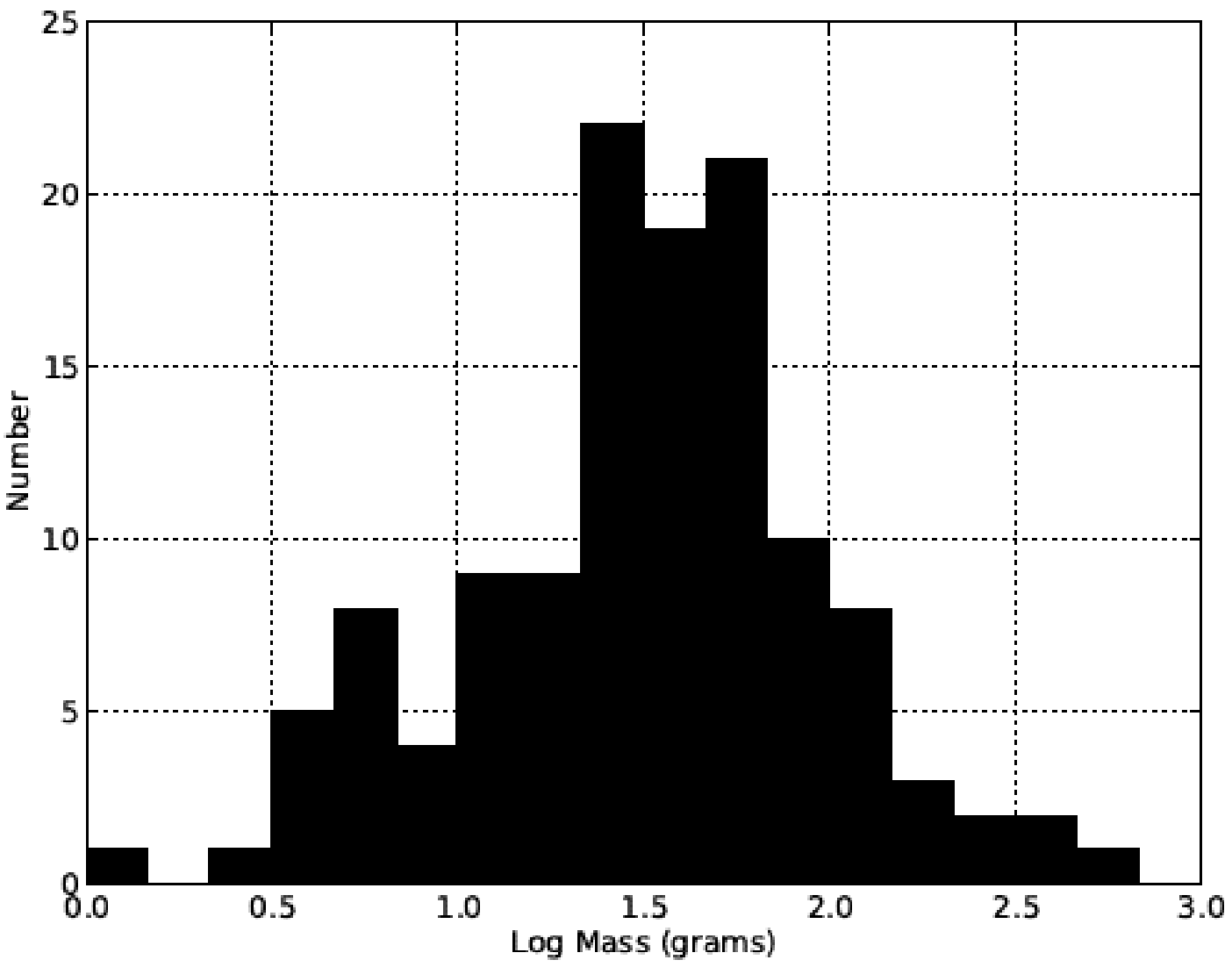

**Fig. 10.** Histogram of impactor masses.

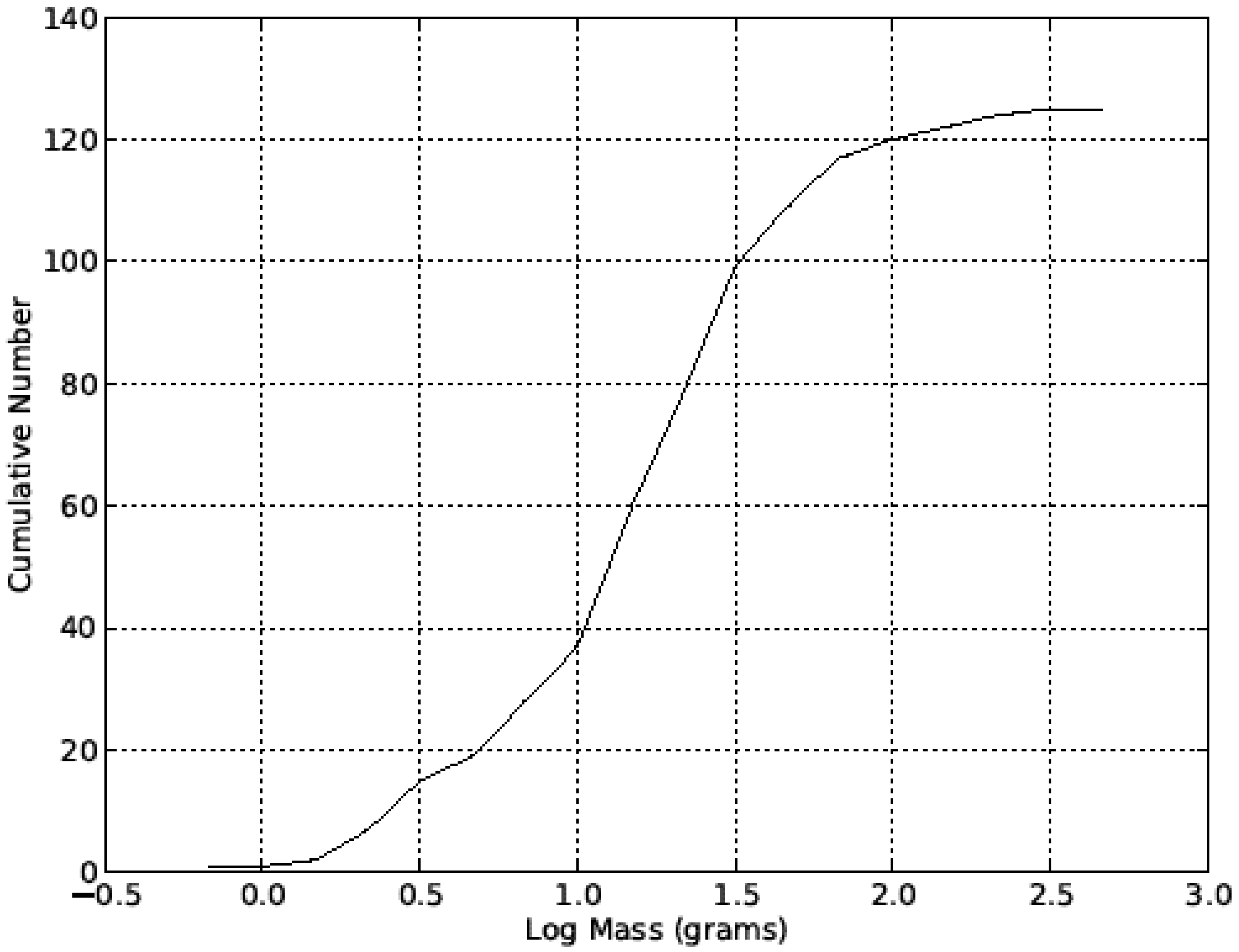

**Fig. 11.** Cumulative histogram of impactor masses showing turnover at a log mass of 1.5 which is approximately the 30 g taken as the completeness limit for this study.

There are 71 meteoroids of the 126 which have masses greater than or equal to this 30 g limit. So the flux is

$$F\,(M \geq 30\text{ g}) = N_{30} \,/\, (\tau A) \qquad (11)$$

where $N_{30}$ = 71 is the number of impact flashes with mass $M$ greater than or equal to 30 g, and the other variables are as before. This calculation produces a flux of $7.00\times10^{-8}$ meteoroids $hr^{-1}$ $km^{-2}$ or $6.14\times10^{-10}$ $m^{-2}$ $yr^{-1}$ for all observed meteoroids with masses greater than 30 g. The Grün et al. (1985) model gives $7.5\times10^{-10}$ $m^{-2}$ $yr^{-1}$ to this mass limit.

## 5. March 17, 2013 Flash and Crater

After the manuscript of this paper was submitted, the Lunar Reconnaissance Orbiter camera team announced they had detected a new crater formed between 12 February 2012 and 28 July 2013 with an 18m diameter measured rim-to-rim (15m inner diameter) located at 20.7135° N, 24.3302° W (Robinson et al., 2014). This crater was likely formed by an impact recorded at MSFC on March 17, 2013 at 03:50:54.312 UTC at that location (Moser et al., 2013). One of the reviewers requested that a short discussion of this impact be added. It is not included in this analysis because it is outside of the 2008 – 2011 time frame discussed here, however it provides a rough check on the photometric calibration and energy estimates. The photometric calibrations described in Appendix A were applied to the video after saturation correction using a Gaussian fit of the flash point spread function, yielding a peak R magnitude of 3.0 ± 0.4 in a 1/30 second video exposure. This corresponds to a luminous energy of $7.1 \times 10^{6}$ J. Note that the full 1/30 second video frame was used in this case because of the extremely long flash duration (approximately 1 second) and the difficulty in Gaussian fitting a point spread function from an interlaced 1/60 second field.

Assuming that the meteoroid impactor was associated with the Virginid Meteor Complex which was observed to be active that night based on observations of fireballs in Earth’s atmosphere by the NASA All Sky Fireball Network (Cooke and Moser, 2011) and the Southern Ontario Meteor Network (Weryk et al., 2008), the speed was 25.6 km/s. Using the velocity-dependent luminous efficiency from equation 8 the impactor kinetic energy was $5.4 \times 10^{9}$ J and the impactor mass was 16 kg. Assuming an impact angle of 56° from horizontal, determined from fireball orbital measurements, a regolith density of 1500 $kg/m^3$, and impactor density between 1800 and 3000 $kg/m^3$, the impact crater inner diameter was estimated to be 9-15 m and 12-20 m rim-to-rim using Holsapple (1993) and Gault (1974) models, consistent with the observed crater size. Therefore the photometric calibration process and luminous efficiency estimates are reasonable.

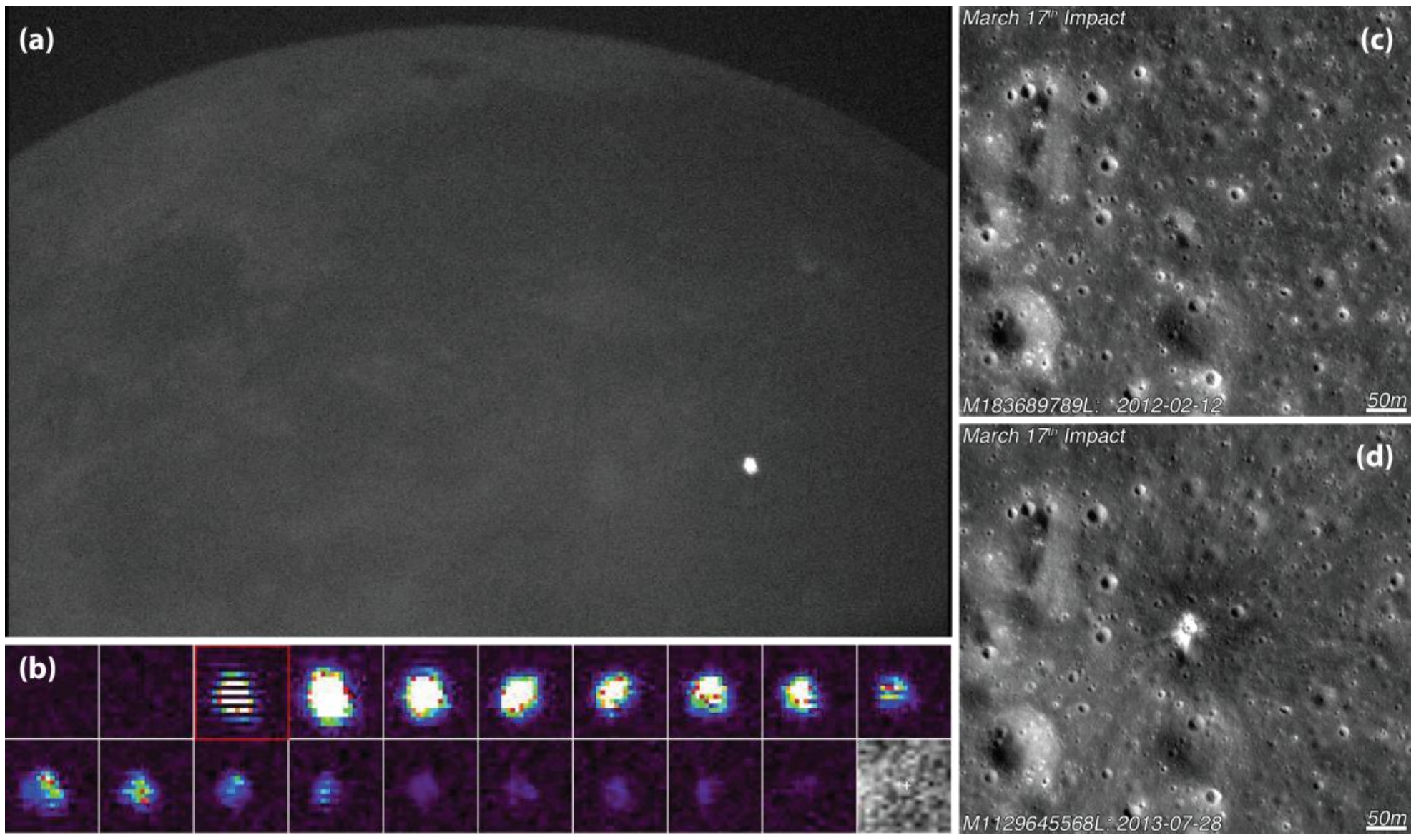

**Fig. 12.** Images of large lunar impact flash observed on March 17, 2013. Panel (a) is the full field-of-view of the brightest frame. Panel (b) is the sequence of successive 1/30 sec video frames zoomed in. Panel (c) is the before and (d) the after image of the impact site captured by the Lunar Reconnaissance Orbiter camera. The rim-to-rim crater diameter is 18 m. Photometric calibration and saturation correction gave an impact energy which resulted in a crater of 12 – 20 m rim-to-rim. See text for details.

## 6. Conclusion

Observations of lunar impacts yield valuable information on the flux of meteoroids in a size range difficult to observe with other means. The flux derived from the impact flashes appears to agree well with the value given by the power law in Figure 9. The agreement of the flux determined using lunar impact monitoring and that predicted by the Grün et al. (1985) model is quite striking.

In spite of limitations in consumer grade video cameras useful photometry is possible using standard CCD calibration techniques. The largest uncertainty in determination of impactor energy is in the luminous efficiency and further work is needed in this area. With sufficient numbers of flash observations it is possible to assign some of the flashes to particular meteor showers, providing a precise impact speed so that calculation of mass from the kinetic energy is possible. Similar observations during the Lunar Atmosphere and Dust Environment Explorer (LADEE) mission and during future lunar seismometer missions (Bouley et al. 2012; Oberst et al. 2012) will provide useful data on impact energies to support the spacecraft science as well as better understand the meteoroid environment.

**Acknowledgements**
The authors gratefully acknowledge the NASA Meteoroid Environment Office, Marshall Space Flight Center Engineering Directorate, the International Space Station Program, Constellation Program, and Space Shuttle Program for financial support for this work. We appreciate the dedication of the following personnel who operated the telescopes and helped analyze the data: Richard Altstatt, Victoria Coffey, Anne Diekmann, Heather Koehler, and Leigh Smith. We also thank Peter Gural for providing the LunarScan software which was essential in performing this work and Rhiannon Blaauw who developed the meteor shower catalog for us. Thanks also to George Varros and Dave Clark for independent confirmation of 3 of the flashes. We owe a huge debt of gratitude to Wesley Swift who helped assemble the observatories and developed the initial photometric analysis software as well as collecting and analyzing some of the data.

The authors also wish to thank Peter Brown and an anonymous reviewer for their thorough review and extremely helpful comments and a third independent reviewer for checking our photometric calibration approach.

## Appendix A. Photometric Calibration and Error Determination

The general approach is outlined in Section 3 and the details are given here. The standard photometric equation (Warner 2006)

$$R = -2.5 \log(S) - k' X + T\,(B\text{-}V) + ZP \qquad \text{(A.1)}$$

was used where $R$ is the Johnson-Cousins R magnitude, $S$ is the sum of the pixel values in the central aperture minus those in the sky aperture, $k'$ is the first-order extinction, $X$ is the airmass, $T$ is the color correction term, $B$-$V$ is the color index of the reference star, and $ZP$ is the zero point for the camera/telescope system. $ZP$ was determined each night using field stars, those passing through the field of view. The values of $k'$ and $T$ were determined using reference stars observed on 2 nights in November 2012 and for the field stars recorded near the Moon each night as described below. The cameras were set for a gamma of 0.45 to extend the dynamic range of the 8 bit systems. Each pixel value was corrected to a linear scale via

$$S = DN^{\,1/0.45} \qquad \text{(A.2)}$$

where $DN$ is the recorded pixel value and $S$ is corrected to a linear scale. The cameras and the settings used are listed in Table A.1

| Setting | Astrovid SellaCam EX | Watec 902H2 Ultimate |
|---|---|---|
| Integration | Sense up – off (none) | Not applicable |
| Gain control | Manual | Manual |
| Shutter control | Off (1/30s frames 1/60 fields) | ELC off (1/30s frames, 1/60 s fields) |
| Gain | Maximum | Adjusted to match StellaCam earthshine |

Table A.1 – Camera settings. These settings were never changed after a camera went into service. The Zero Point determined each night (equation A.1) accounts for any drift in gain.

Note that six of the brightest flashes were saturated which means they exceeded the dynamic range of the camera. No corrections have been made for these so their calculated energies are actually a lower limit. Those flashes which were saturated are indicated in bold in Tables 1 and 3 and are flagged in Figures 5 and 6. Inspection of the point spread functions showed that even the brightest flashes had only minimal saturation (compared to the March 17, 2013 event) so the actual energy of the flashes would be low by only a factor of 2 or less. This affects some of the most energetic points which fall below the power law on Figure 9 but does not explain the downward trend with energy.

*A.1 Color Correction*

Since maximum sensitivity was needed no filters were used on the Sony EX HAD-based cameras. Their broad spectral response (Figure A.1) necessitated the color correction term, *T*, which was determined by observing 19 reference stars with airmasses less than 1.15. This color correction term empirically corrects for the throughput of the telescope optics and the response of the camera. The peak of the CCD response is close to that of the R filter so stellar magnitudes in that filter were used for calibration. Each of the stars, although they were not photometric standards, had measured R magnitudes and were not variable stars according to SIMBAD (http://simbad.u-strasbg.fr/simbad/). An additional 12 reference stars were measured at airmasses of 1.2 – 2.8 in order to determine an extinction coefficient *k'* and for use in the characterization of errors due to scintillation described in Section A.2.

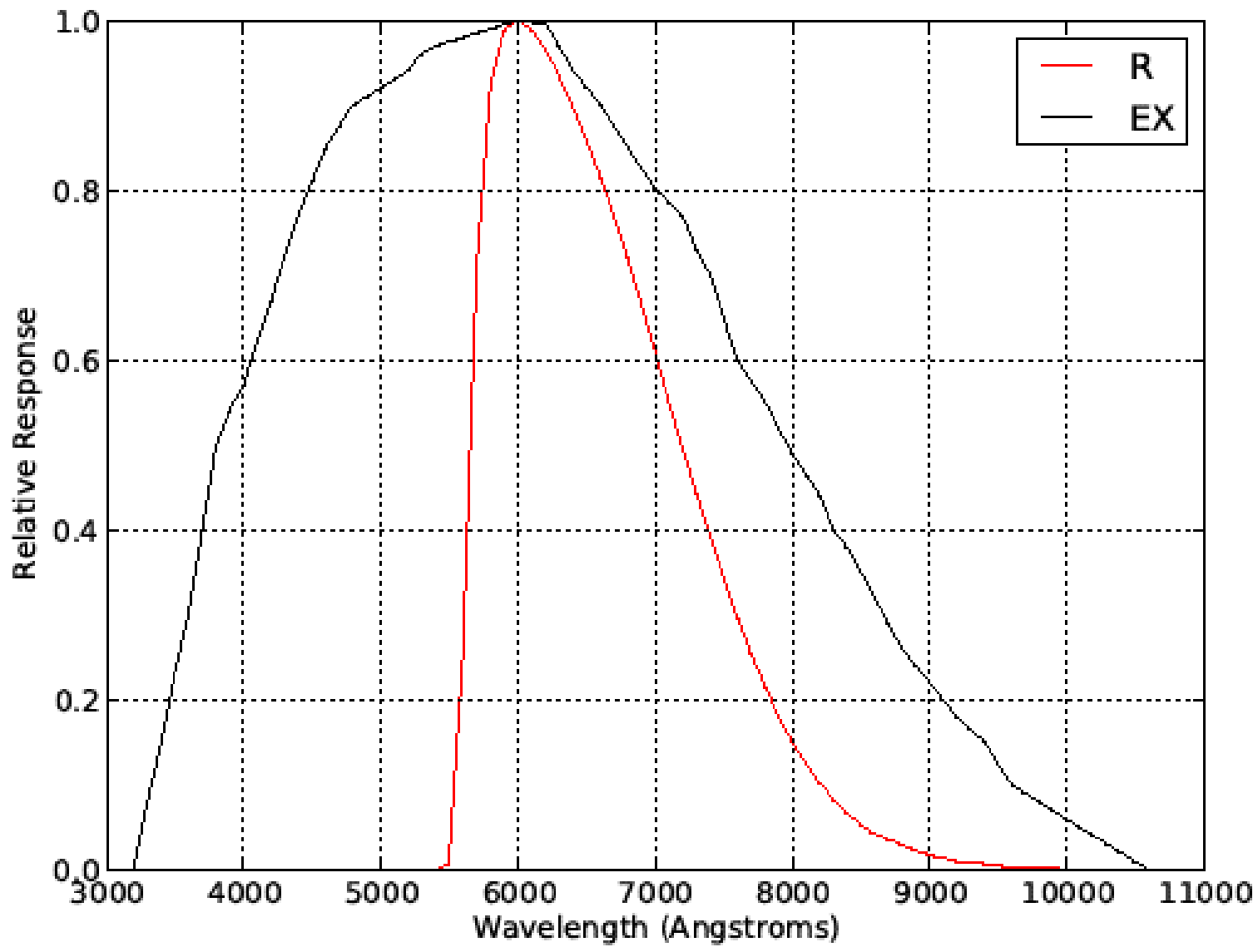

**Fig. A.1.** Johnson-Cousins R filter (red) and camera response curve (black).

Stars passing through the field during lunar observations were used to determine *k'* (if there were 3 or more stars for the night) and *ZP*. R magnitudes were generally not available for these field stars so a fit of *B-V* vs. *V-R* for the Landolt standards (Landolt 1992) was computed. Only standard stars bluer than *B-V* = 1.2 were used since the *B-V* vs *V-R* relationship diverges for redder stars. The least squares fit (Figure A.2) gave

$$R = V - 0.019 - 0.562\,(B-V) \qquad \text{(A.3)}$$

The standard deviation of this fit is 0.04 magnitudes.

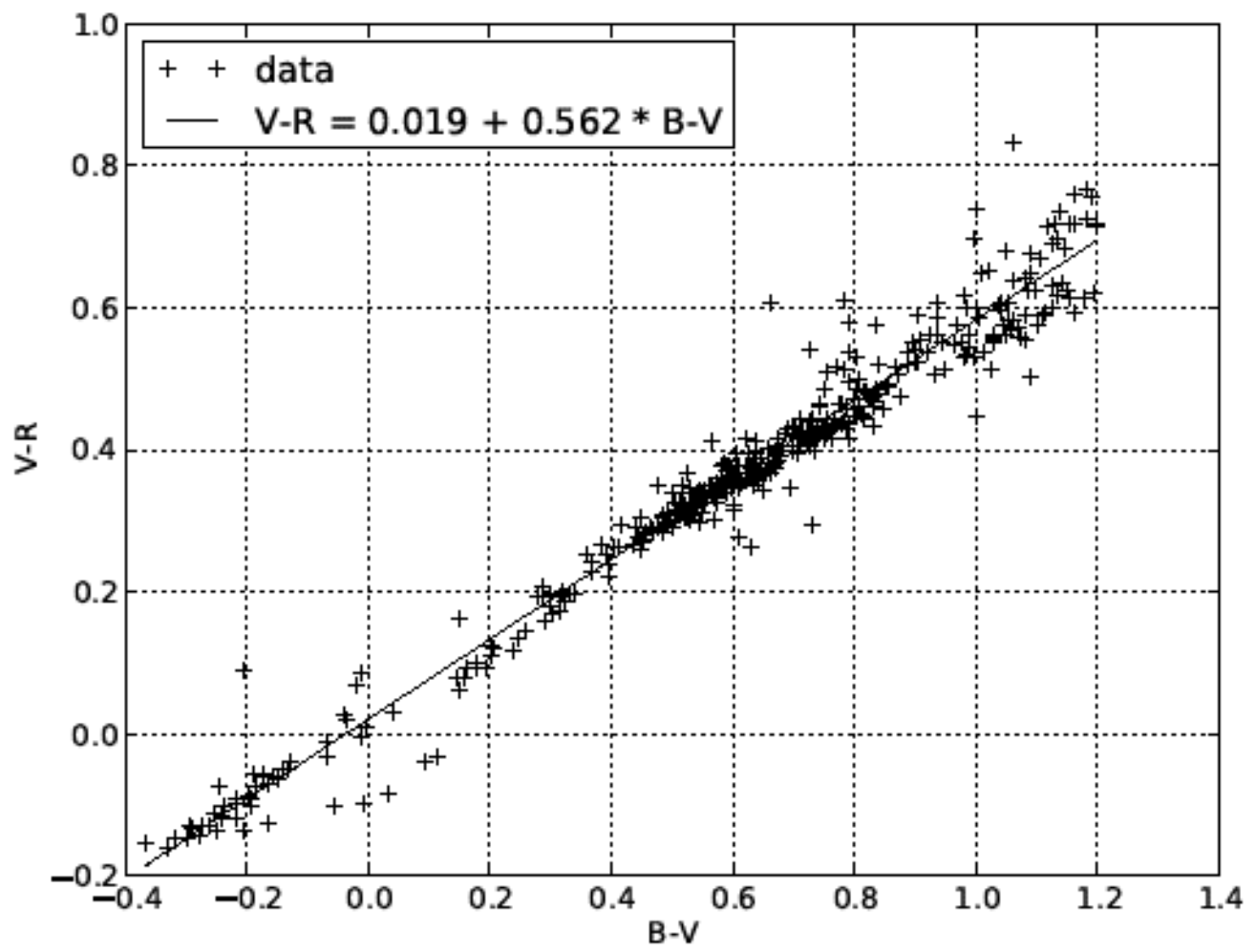

**Fig. A.2.** Landolt photometric standard stars *V-R* versus *B-V* with linear fit.

An average *k'* for the ensemble of all 344 field star measurements was 0.311 which is almost identical to the *k'* = 0.319 determined for the group of reference stars. Magnitudes for the flashes were computed using *k'* for the night when there were 3 or more stars and clouds didn't affect the stellar observations. Otherwise, the *k'* determined from the reference stars was used. Hardie (1962) suggests using average extinction where an insufficient number of extinction measures are available. For extremely humid, hazy nights *k'* was as high as 0.6 magnitudes per airmass and the extinction coefficient determined from passing field stars was used for those nights.

The *B-V* color index for the impact flashes is unknown, so the *T* color correction term cannot be used in the flash analysis. Instead, a blackbody temperature of 2800K was adopted from the modeling of Nemtchinov et al. (1998). Convolutions of the spectral responses of the R filter (Bessell et al. 1998) and the EX HAD chip (SONY n.d.) were compared to determine a (*R-EX*) correction term to replace the *T* (*B-V*) term in the standard photometric equation. Figure A.3 shows the EX HAD to R correction factor versus blackbody temperature. This correction term varies slowly with temperature between 2500 and 4500K. Ernst and Schultz (2005) observed color temperatures ranging from 3000 – 4500K in Ames Vertical Gun Range hypervelocity impact tests. So this correction factor of *R-EX* = -0.66 should cover the range of peak temperatures of impact flashes. As a check on this *R-EX* value a *B-V* of 2.4 for a 2800 K star was computed using the relation in Sekiguchi and Fukujita (2000). A star this cool would not be a perfect blackbody due to absorption in its atmosphere, however this *B-V* gives a *T (B-V)* of -0.7, reasonably close to the *R-EX* value.

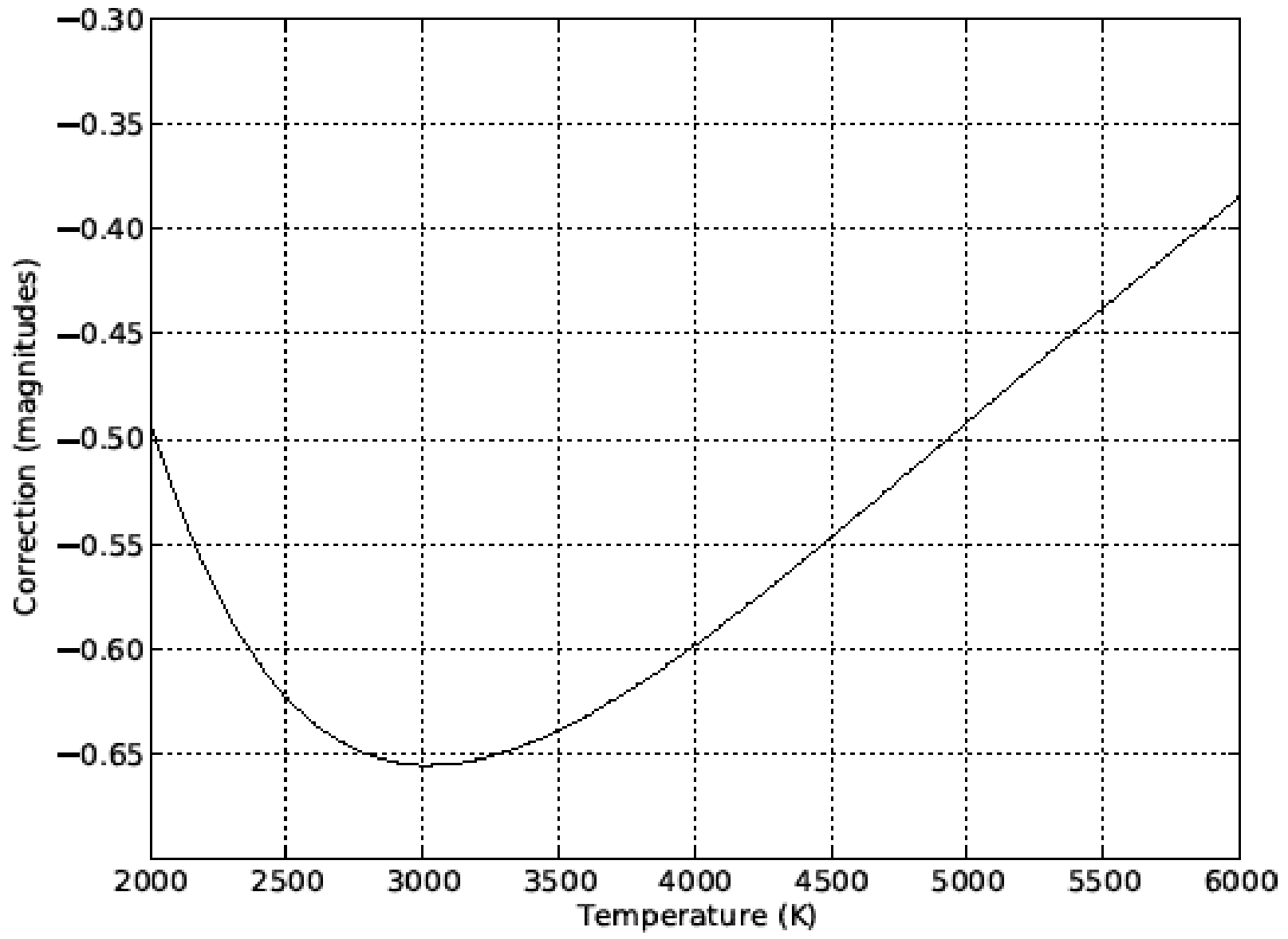

**Fig. A.3.** Magnitude correction term from Johnson-Cousins R and Sony EX HAD CCD response for a range of blackbody temperatures.

*A.2 Photometric Errors*

Photometry with consumer-grade, 8 bit, broadband video cameras is not precise but standard photometric techniques allow useful data to be obtained. To determine the reliability of these measurements several sources of error were addressed and characterized. The lack of photometric resolution for the 8 bit per pixel system was one source. The broad spectral response and the adjustment to the R filter passband are addressed in the previous section. Although the rise time of the flashes is extremely short, it is possible for the video exposure to start during this rise time. Occasionally a video field shows a faint flash image followed by a brighter one. This analysis uses the brightest field and, for the 72 flashes where magnitudes were determined with 2 telescopes, the brighter of the 2 measurements. Since the camera exposures were not synchronized, the brightest measurement is more likely to include all of the light from the initial flash produced when the impactor struck the lunar surface (as opposed to the signature from cooling ejecta).

Scintillation caused by atmospheric density variations along the line of sight is especially significant for these relatively short exposures. This effect was investigated by measuring the field-to-field variation in instrumental magnitude, -2.5 log ($S$), over 30 s intervals of the reference stars observed at a range of airmasses. A fit of the 1 sigma variation gives

$$\sigma_{scint} = 0.0056 + 0.076\, X \qquad \text{(A.4)}$$

where $X$ is the airmass of the observation. Figure A.4 shows the fit determined from the reference star observations.

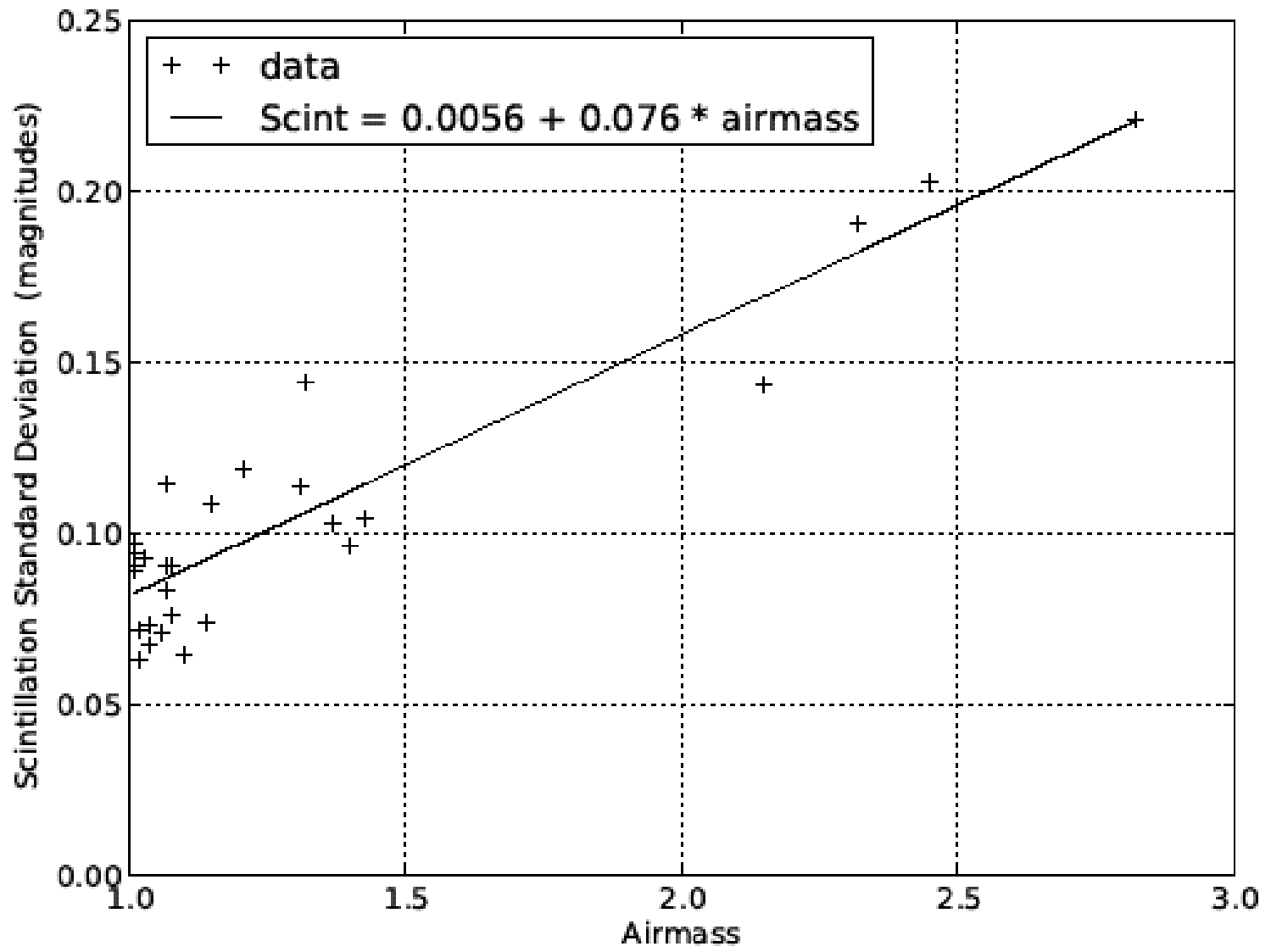

**Fig. A.4.** Field-to-field standard deviation of instrumental magnitude versus airmass to determine photometric error due to atmospheric scintillation.

Residuals of the fit indicates a 1 sigma error ($\sigma_{fit}$) of 0.2 magnitudes. Combining this error with the scintillation error

$$\sigma^2 = \sigma_{scint}^2 + \sigma_{fit}^2 \quad \text{(A.5)}$$

gives the photometric uncertainties in R magnitude shown in Table 1. These uncertainties are propagated to the error bars in the kinetic energy and mass values shown in Figures 5 and 6.

**Table 1**
Impact flashes observed from September 2006-August 2011 during the NASA Lunar Impact Monitoring Program. For each flash, numbered consecutively, the date, time, and solar longitude of the observation is given. N indicates the number of telescopes that observed the flash. The latitude and longitude of the impact flash in selenographic coordinates are given in columns 6 and 7. Airmass at the time of the observation, the peak R magnitude of the flash, and the peak luminous energy are shown in the last 3 columns. Six flashes which exceeded the dynamic range of the camera as described in Appendix A are indicated in bold. The energies for those are lower limits.

| Flash No. | Date (UT) | Time (UT) | Solar Long (°) | N | Lunar Lat (°) | Lunar Long (°) | Air Mass | Peak R Mag | Peak Lum. Energy (J) |
|---|---|---|---|---|---|---|---|---|---|
| 1 | 16 Sep 2006 | 09:52:54 | 173.3082 | 2 | -32.0 | 57.0 | 1.40 | 8.5 ± 0.2 | $2.14\times10^4$ |
| 2 | 28 Sep 2006 | 00:42:57 | 184.6750 | 2 | -0.5 | -30.8 | 4.00 | 8.1 ± 0.4 | $3.14\times10^4$ |
| 3 | 29 Oct 2006 | 01:18:43 | 215.3898 | 2 | 1.7 | -45.7 | 2.61 | 7.6 ± 0.3 | $4.98\times10^4$ |
| 4 | 17 Nov 2006 | 10:46:26 | 234.8499 | 2 | 42.6 | 76.7 | 3.38 | 8.9 ± 0.3 | $1.56\times10^{4}$ |
| 5 | 17 Nov 2006 | 10:56:33 | 234.8570 | 2 | 36.1 | 80.3 | 3.04 | 7.0 ± 0.3 | $8.58\times10^{4}$ |
| 6 | 17 Nov 2006 | 11:02:29 | 234.8611 | 2 | 5.0 | 85.5 | 2.91 | 7.6 ± 0.3 | $5.17\times10^{4}$ |
| 7 | 17 Nov 2006 | 11:09:10 | 234.8658 | 2 | -10.0 | 66.8 | 2.75 | 8.6 ± 0.3 | $1.93\times10^{4}$ |
| 8 | 24 Nov 2006 | 23:58:12 | 242.4758 | 2 | -37.8 | -28.7 | 3.40 | 7.8 ± 0.3 | $4.11\times10^{4}$ |
| 9 | 25 Nov 2006 | 00:55:54 | 242.5163 | 2 | 1.3 | -80.8 | 6.18 | 7.9 ± 0.5 | $3.78\times10^{4}$ |
| 10 | 26 Nov 2006 | 00:59:16 | 243.5303 | 2 | 41.3 | -21.2 | 3.05 | 7.5 ± 0.3 | $5.67\times10^{4}$ |
| 11 | 26 Nov 2006 | 01:28:43 | 243.5511 | 2 | 18.2 | -32.0 | 3.83 | 8.7 ± 0.4 | $1.78\times10^{4}$ |
| 12 | 26 Nov 2006 | 01:30:29 | 243.5523 | 2 | 28.3 | -32.5 | 3.90 | 8.7 ± 0.4 | $1.86\times10^{4}$ |
| 13 | 14 Dec 2006 | 08:46:02 | 262.1186 | 2 | 14.3 | 79.9 | 3.04 | 9.3 ± 0.3 | $1.07\times10^{4}$ |
| 14 | 14 Dec 2006 | 08:50:35 | 262.1219 | 2 | 12.3 | 46.4 | 2.91 | 8.2 ± 0.3 | $2.84\times10^{4}$ |
| 15 | 14 Dec 2006 | 08:51:20 | 262.1224 | 2 | -11.0 | 51.4 | 2.91 | 8.8 ± 0.3 | $1.62\times10^{4}$ |
| 16 | 14 Dec 2006 | 08:56:42 | 262.1262 | 2 | -5.3 | 84.0 | 2.77 | 8.4 ± 0.3 | $2.30\times10^{4}$ |
| 17 | 14 Dec 2006 | 09:00:21 | 262.1288 | 2 | 40.0 | 39.2 | 2.70 | 8.4 ± 0.3 | $2.34\times10^{4}$ |
| 18 | 14 Dec 2006 | 09:03:32 | 262.1310 | 2 | 22.2 | 61.2 | 2.62 | 9.9 ± 0.3 | $5.83\times10^{3}$ |
| 19 | 15 Dec 2006 | 09:15:14 | 263.1566 | 2 | 37.3 | 84.8 | 4.21 | 8.5 ± 0.4 | $2.22\times10^{4}$ |
| 20 | 15 Dec 2006 | 09:17:39 | 263.1583 | 2 | 26.6 | 60.2 | 4.06 | 7.4 ± 0.4 | $6.05\times10^{4}$ |
| 21 | 23 Feb 2007 | 00:47:45 | 333.9213 | 2 | -6.8 | -12.8 | 1.19 | 7.9 ± 0.2 | $3.61\times10^{4}$ |
| 22 | 23 Feb 2007 | 04:02:44 | 334.0576 | 2 | 21.0 | -33.9 | 3.04 | 9.0 ± 0.3 | $1.32\times10^{4}$ |
| 23 | 22 Apr 2007 | 01:15:05 | 31.4533 | 2 | -31.5 | -19.8 | 1.31 | 9.3 ± 0.2 | $9.95\times10^{3}$ |
| 24 | 22 Apr 2007 | 01:15:41 | 31.4538 | 2 | -9.8 | -19.2 | 1.31 | 9.3 ± 0.2 | $1.02\times10^{4}$ |
| 25 | 22 Apr 2007 | 01:38:31 | 31.4692 | 3[a] | 26.4 | -19.0 | 1.41 | 7.6 ± 0.2 | $4.94\times10^{4}$ |
| 26 | 22 Apr 2007 | 03:12:21 | 31.5328 | 3[b] | 24.0 | -19.2 | 2.19 | 6.7 ± 0.3 | $1.11\times10^{5}$ |
| 27 | 22 Apr 2007 | 03:52:37 | 31.5601 | 2 | -4.8 | -17.6 | 2.95 | 8.6 ± 0.3 | $2.06\times10^{4}$ |
| 28 | 22 Apr 2007 | 04:22:27 | 31.5803 | 2 | 30.0 | -43.8 | 4.02 | 8.2 ± 0.4 | $2.95\times10^{4}$ |
| 29 | 23 Apr 2007 | 01:15:55 | 32.4294 | 2 | 6.8 | -48.2 | 1.15 | 8.3 ± 0.2 | $2.64\times10^{4}$ |
| 30 | 23 Apr 2007 | 02:23:21 | 32.4751 | 2 | 5.0 | -18.6 | 1.36 | 8.8 ± 0.2 | $1.61\times10^{4}$ |
| 31 | 23 Apr 2007 | 03:01:10 | 32.5007 | 2 | -24.4 | -70.0 | 1.56 | 8.6 ± 0.2 | $1.97\times10^{4}$ |
| 32 | 23 Apr 2007 | 04:08:49 | 32.5465 | 2 | 26.2 | -27.5 | 2.24 | 8.0 ± 0.3 | $3.32\times10^{4}$ |
| 33 | 23 Apr 2007 | 04:38:22 | 32.5665 | 2 | 12.0 | -39.5 | 2.78 | 8.6 ± 0.3 | $2.06\times10^{4}$ |
| 34 | 23 Apr 2007 | 04:40:46 | 32.5682 | 2 | -11.0 | -38.0 | 2.85 | 9.3 ± 0.3 | $1.02\times10^{4}$ |
| 35 | 23 Apr 2007 | 04:42:35 | 32.5694 | 2 | -10.2 | -28.2 | 2.88 | 7.0 ± 0.3 | $8.43\times10^{4}$ |
| 36 | 23 Apr 2007 | 04:59:58 | 32.5811 | 2 | 30.0 | -60.0 | 3.42 | 7.7 ± 0.3 | $4.63\times10^{4}$ |
| 37 | 21 May 2007 | 02:50:53 | 59.6059 | 2 | 0.0 | -26.4 | 2.30 | 7.8 ± 0.3 | $4.00\times10^{4}$ |
| 38 | 21 May 2007 | 03:10:07 | 59.6188 | 2 | 7.3 | -68.3 | 2.64 | 9.0 ± 0.3 | $1.34\times10^{4}$ |
| 39 | 09 Aug 2007 | 09:10:50 | 136.3210 | 2 | 10.4 | 48.7 | 2.85 | 7.3 ± 0.3 | $6.45\times10^{4}$ |
| 40 | 06 Oct 2007 | 08:42:52 | 192.6138 | 3 | 24.3 | 75.6 | 3.68 | 8.6 ± 0.4 | $2.02\times10^{4}$ |
| 41 | 16 Nov 2007 | 00:11:21 | 233.1410 | 2 | -37.0 | -55.6 | 2.20 | 8.9 ± 0.3 | $1.44\times10^{4}$ |
| **42** | **16 Nov 2007** | **00:27:09** | **233.1520** | **2** | **32.5** | **-58.2** | **2.33** | **7.1 ± 0.3** | $\mathbf{8.12\times10^{4}}$ |
| 43 | 03 Jan 2008 | 10:25:42 | 282.2998 | 2 | 26.3 | 89.3 | 3.47 | 9.1 ± 0.3 | $1.20\times10^{4}$ |
| 44 | 04 Jan 2008 | 10:58:26 | 283.3426 | 3 | 39.9 | 38.0 | 4.71 | 8.2 ± 0.4 | $2.92\times10^{4}$ |
| **45** | **04 Jan 2008** | **11:42:39** | **283.3739** | **3** | **39.6** | **72.0** | **3.21** | **7.1 ± 0.3** | $\mathbf{7.54\times10^{4}}$ |
| 46 | 04 Jan 2008 | 11:48:36 | 283.3781 | 3 | 34.0 | 64.0 | 3.08 | 8.3 ± 0.3 | $2.66\times10^{4}$ |
| 47 | 14 Jan 2008 | 00:22:26 | 293.0868 | 3 | -24.4 | -49.0 | 1.40 | 9.7 ± 0.2 | $7.48\times10^{3}$ |
| 48 | 11 Feb 2008 | 01:09:27 | 321.5676 | 3 | -17.8 | -27.2 | 2.20 | 8.2 ± 0.3 | $2.89\times10^{4}$ |
| 49 | 12 Feb 2008 | 00:24:44 | 322.5482 | 3 | -12.0 | -76.5 | 1.33 | 8.6 ± 0.2 | $2.02\times10^{4}$ |
| **50** | **12 Mar 2008** | **00:40:42** | **351.7098** | **3** | **-24.8** | **-23.2** | **1.37** | **6.8 ± 0.2** | $\mathbf{1.03\times10^{5}}$ |
| 51 | 12 Mar 2008 | 01:13:31 | 351.7325 | 3 | -7.0 | -49.5 | 1.54 | 9.1 ± 0.2 | $1.20\times10^{4}$ |
| 52 | 12 Mar 2008 | 02:03:07 | 351.7669 | 3 | -6.4 | -56.4 | 1.96 | 9.3 ± 0.3 | $1.05\times10^{4}$ |
| 53 | 13 Mar 2008 | 01:38:48 | 352.7479 | 3 | -2.5 | -48.0 | 1.33 | 10.4 ± 0.2 | $3.68\times10^{3}$ |
| **54** | **13 Mar 2008** | **02:04:22** | **352.7656** | **4[b]** | **-23.0** | **-77.0** | **1.45** | **7.7 ± 0.2** | $\mathbf{4.59\times10^{4}}$ |
| 55 | 09 Apr 2008 | 02:16:38 | 19.4998 | 2 | -16.4 | -32.4 | 3.68 | 6.1 ± 0.4 | $2.02\times10^{5}$ |
| 56 | 10 Apr 2008 | 01:15:25 | 20.4403 | 2 | -5.0 | -66.0 | 1.54 | 8.4 ± 0.2 | $2.34\times10^{4}$ |
| 57 | 07 Jun 2008 | 02:27:23 | 76.6099 | 3 | 22.8 | -78.5 | 3.03 | 6.6 ± 0.3 | $1.25\times10^{5}$ |
| 58 | 07 Jun 2008 | 03:31:31 | 76.6525 | 2 | 3.4 | -11.8 | 7.77 | 7.9 ± 0.7 | $3.61\times10^{4}$ |
| 59 | 27 Jun 2008 | 09:31:25 | 95.9845 | 3 | 0.3 | 78.1 | 1.42 | 8.7 ± 0.2 | $1.84\times10^{4}$ |
| 60 | 28 Jul 2008 | 08:35:37 | 125.5209 | 3 | 1.8 | 72.7 | 2.50 | 10.1 ± 0.3 | $4.80\times10^{3}$ |
| 61 | 28 Jul 2008 | 08:47:07 | 125.5285 | 3 | 27.8 | 24.4 | 2.31 | 8.8 ± 0.3 | $1.61\times10^{4}$ |
| 62 | 28 Jul 2008 | 09:05:16 | 125.5405 | 3 | 2.6 | 70.0 | 2.05 | 8.5 ± 0.3 | $2.22\times10^{4}$ |
| 63 | 29 Jul 2008 | 09:43:11 | 126.5219 | 3 | 22.2 | 64.0 | 2.42 | 7.8 ± 0.3 | $3.96\times10^{4}$ |
| 64 | 23 Sep 2008 | 10:14:33 | 180.6341 | 2 | 3.0 | 77.0 | 1.22 | 8.9 ± 0.2 | $1.46\times10^{4}$ |
| 65 | 24 Sep 2008 | 08:46:41 | 181.5537 | 3 | -25.0 | 65.0 | 2.33 | 6.3 ± 0.3 | $1.58\times10^{5}$ |
| 66 | 22 Oct 2008 | 07:25:10 | 209.1435 | 2 | -15.0 | 71.0 | 2.75 | 8.0 ± 0.3 | $3.45\times10^{4}$ |
| 67 | 22 Oct 2008 | 07:51:31 | 209.1617 | 2 | 19.5 | 94.0 | 2.25 | 8.5 ± 0.3 | $2.26\times10^{4}$ |
| 68 | 22 Oct 2008 | 10:03:14 | 209.2528 | 2 | 33.0 | 75.0 | 1.27 | 9.0 ± 0.2 | $1.40\times10^{4}$ |
| 69 | 22 Oct 2008 | 10:30:21 | 209.2715 | 3 | 5.5 | 79.0 | 1.19 | 8.9 ± 0.2 | $1.44\times10^{4}$ |
| 70 | 02 Nov 2008 | 23:48:40 | 220.8110 | 3 | -11.0 | -82.5 | 2.90 | 8.8 ± 0.3 | $1.58\times10^{4}$ |
| 71 | 03 Nov 2008 | 00:11:06 | 220.8266 | 3 | 12.5 | -70.5 | 3.31 | 7.7 ± 0.3 | $4.38\times10^{4}$ |
| 72 | 03 Nov 2008 | 23:59:24 | 221.8204 | 3 | -5.0 | -65.0 | 2.30 | 8.9 ± 0.3 | $1.45\times10^{4}$ |
| 73 | 04 Nov 2008 | 00:04:06 | 221.8236 | 3 | 19.0 | -52.0 | 2.34 | 8.5 ± 0.3 | $2.20\times10^{4}$ |
| 74 | 04 Nov 2008 | 01:10:01 | 221.8695 | 3 | -24.5 | -21.5 | 3.22 | 8.5 ± 0.3 | $2.16\times10^{4}$ |
| 75 | 04 Nov 2008 | 01:39:04 | 221.8897 | 3 | -1.5 | -62.0 | 4.08 | 7.5 ± 0.4 | $5.62\times10^{4}$ |
| 76 | 05 Nov 2008 | 00:38:38 | 222.8500 | 3 | -17.0 | -66.5 | 2.04 | 7.5 ± 0.3 | $5.32\times10^{4}$ |
| 77 | 05 Nov 2008 | 00:53:58 | 222.8607 | 3 | -23.0 | -18.0 | 2.13 | 9.0 ± 0.3 | $1.37\times10^{4}$ |
| 78 | 05 Nov 2008 | 02:05:08 | 222.9102 | 3 | 30.0 | -65.0 | 2.99 | 7.2 ± 0.3 | $7.48\times10^{4}$ |
| 79 | 05 Nov 2008 | 02:09:45 | 222.9134 | 3 | -35.0 | -43.0 | 3.10 | 9.4 ± 0.3 | $9.41\times10^{3}$ |
| 80 | 05 Nov 2008 | 02:32:47 | 222.9295 | 3 | -1.0 | -33.0 | 3.74 | 8.0 ± 0.4 | $3.58\times10^{4}$ |
| 81 | 20 Nov 2008 | 11:03:06 | 238.3758 | 2 | -5.0 | 52.0 | 1.17 | 8.7 ± 0.2 | $1.79\times10^{4}$ |
| 82 | 22 Nov 2008 | 09:41:24 | 240.3387 | 2 | 1.5 | 84.0 | 2.55 | 7.8 ± 0.3 | $4.11\times10^{4}$ |
| 83 | 23 Nov 2008 | 10:48:24 | 241.3966 | 3 | -1.0 | 86.0 | 2.52 | 8.7 ± 0.3 | $1.74\times10^{4}$ |
| 84 | 23 Nov 2008 | 11:15:53 | 241.4159 | 3 | -21.0 | 46.0 | 2.13 | 9.1 ± 0.3 | $1.25\times10^{4}$ |
| 85 | 03 Dec 2008 | 00:30:58 | 251.0915 | 3 | -18.0 | -63.0 | 2.47 | 10.1 ± 0.3 | $5.08\times10^{3}$ |
| 86 | 03 Dec 2008 | 02:09:04 | 251.1606 | 2 | -4.5 | -68.5 | 6.45 | 7.9 ± 0.6 | $3.68\times10^{4}$ |
| 87 | 01 Feb 2009 | 01:40:26 | 312.2038 | 3 | -18.0 | -66.0 | 1.71 | 9.4 ± 0.2 | $9.67\times10^{3}$ |
| 88 | 01 Feb 2009 | 02:04:46 | 312.2210 | 3 | 25.0 | -70.0 | 1.94 | 8.9 ± 0.3 | $1.49\times10^{4}$ |
| 89 | 02 Feb 2009 | 02:45:43 | 313.2647 | 3 | -21.0 | -55.0 | 1.73 | 8.8 ± 0.2 | $1.62\times10^{4}$ |
| 90 | 03 Mar 2009 | 02:51:43 | 342.5371 | 3 | -32.5 | -35.0 | 1.77 | 9.2 ± 0.2 | $1.09\times10^{4}$ |
| 91 | 03 Mar 2009 | 04:02:49 | 342.5866 | 3 | -23.0 | -84.0 | 2.80 | 8.7 ± 0.3 | $1.79\times10^{4}$ |
| 92 | 03 Mar 2009 | 04:27:49 | 342.6040 | 3 | 9.0 | -73.0 | 3.57 | 8.0 ± 0.3 | $3.48\times10^{4}$ |
| 93 | 30 Mar 2009 | 01:43:11 | 9.3700 | 3 | 27.5 | -68.0 | 2.56 | 8.3 ± 0.3 | $2.50\times10^{4}$ |
| 94 | 30 May 2009 | 03:52:11 | 68.7580 | 3 | 11.0 | -15.0 | 2.74 | 6.8 ± 0.3 | $1.06\times10^{5}$ |
| 95 | 19 Jun 2009 | 09:00:07 | 88.0881 | 3 | -32.0 | 54.0 | 3.08 | 7.5 ± 0.3 | $5.67\times10^{4}$ |
| 96 | 26 Jun 2009 | 02:04:07 | 94.4932 | 2 | 16.0 | -56.0 | 3.12 | 7.6 ± 0.3 | $4.89\times10^{4}$ |
| 97 | 24 Oct 2009 | 23:57:36 | 211.5755 | 3 | 11.5 | -80.0 | 1.90 | 8.7 ± 0.3 | $1.79\times10^{4}$ |
| 98 | 25 Oct 2009 | 00:14:24 | 211.5871 | 2 | 15.5 | -65.0 | 1.96 | 10.4 ± 0.3 | $3.89\times10^{3}$ |
| 99 | 25 Oct 2009 | 01:20:00 | 211.6325 | 3 | -24.5 | -60.0 | 2.37 | 8.8 ± 0.3 | $1.59\times10^{4}$ |
| 100 | 25 Oct 2009 | 01:52:04 | 211.6547 | 3 | 18.0 | -70.0 | 2.78 | 9.5 ± 0.3 | $8.90\times10^{3}$ |
| 101 | 25 Oct 2009 | 01:55:07 | 211.6568 | 3 | -24.0 | -38.5 | 2.83 | 8.6 ± 0.3 | $1.93\times10^{4}$ |
| 102 | 25 Oct 2009 | 01:58:10 | 211.6589 | 2 | 35.0 | -22.5 | 2.89 | 8.0 ± 0.3 | $3.35\times10^{4}$ |
| 103 | 25 Oct 2009 | 02:37:58 | 211.6865 | 3 | 3.0 | -41.5 | 3.93 | 8.7 ± 0.4 | $1.78\times10^{4}$ |
| 104 | 25 Oct 2009 | 02:38:08 | 211.6866 | 3 | 4.5 | -56.0 | 3.95 | 8.5 ± 0.4 | $2.16\times10^{4}$ |
| 105 | 25 Oct 2009 | 02:53:59 | 211.6976 | 3 | 29.0 | -69.5 | 4.69 | 8.0 ± 0.4 | $3.48\times10^{4}$ |
| 106 | 12 Nov 2009 | 10:10:23 | 230.0223 | 3 | 16.0 | 77.0 | 2.34 | 8.4 ± 0.3 | $2.34\times10^{4}$ |
| 107 | 13 Nov 2009 | 10:15:45 | 231.0324 | 3 | 8.0 | 53.0 | 4.16 | 7.9 ± 0.4 | $3.78\times10^{4}$ |
| 108 | 10 Dec 2009 | 09:41:54 | 258.3249 | 2 | -7.5 | 87.0 | 1.99 | 8.5 ± 0.3 | $2.24\times10^{4}$ |
| 109 | 21 Dec 2009 | 00:12:38 | 269.1158 | 3 | 28.5 | -35.0 | 2.32 | 7.6 ± 0.3 | $4.76\times10^{4}$ |
| 110 | 21 Dec 2009 | 00:33:46 | 269.1308 | 2 | -32.0 | -36.0 | 2.65 | 8.5 ± 0.3 | $2.08\times10^{4}$ |
| 111 | 19 Feb 2010 | 00:45:39 | 330.1181 | 2 | -36.0 | -40.0 | 1.60 | 8.9 ± 0.2 | $1.56\times10^{4}$ |
| 112 | 19 Feb 2010 | 01:15:11 | 330.1388 | 2 | -11.5 | -73.5 | 1.84 | 8.4 ± 0.2 | $2.41\times10^{4}$ |
| 113 | 21 Apr 2010 | 02:48:03 | 30.7640 | 3 | 13.0 | -31.0 | 1.44 | 8.9 ± 0.2 | $1.49\times10^{4}$ |
| 114 | 18 May 2010 | 01:38:31 | 56.8896 | 3 | -0.5 | -36.5 | 1.82 | 9.0 ± 0.2 | $1.40\times10^{4}$ |
| 115 | 18 May 2010 | 01:56:33 | 56.9017 | 2 | 12.0 | -62.5 | 2.00 | 9.7 ± 0.3 | $7.07\times10^{3}$ |
| 116 | 18 May 2010 | 02:31:10 | 56.9248 | 3 | 16.0 | -52.0 | 2.54 | 7.5 ± 0.3 | $5.22\times10^{4}$ |
| **117** | **08 Jul 2010** | **08:48:56** | **105.9568** | **2** | **-36.0** | **20.0** | **3.48** | **5.1 ± 0.3** | **$5.08\times10^{5}$** |
| 118 | 02 Sep 2010 | 06:54:16 | 159.5649 | 3 | -6.5 | 36.0 | 2.45 | 8.3 ± 0.3 | $2.50\times10^{4}$ |
| 119 | 04 Oct 2010 | 09:27:01 | 190.9047 | 2 | -5.5 | 58.0 | 3.40 | 8.2 ± 0.3 | $2.92\times10^{4}$ |
| 120 | 09 Jan 2011 | 01:17:55 | 288.2609 | 3 | -37.0 | -55.0 | 2.63 | 8.4 ± 0.3 | $2.39\times10^{4}$ |
| 121 | 26 Feb 2011 | 09:39:29 | 337.2799 | 3 | 15.0 | 69.5 | 3.89 | 8.9 ± 0.4 | $1.48\times10^{4}$ |
| 122 | 26 Feb 2011 | 10:38:26 | 337.3211 | 3 | 7.0 | 48.0 | 2.61 | 8.9 ± 0.3 | $1.52\times10^{4}$ |
| 123 | 08 Apr 2011 | 01:32:18 | 17.7286 | 3 | -2.0 | -46.0 | 1.93 | 7.4 ± 0.3 | $6.16\times10^{4}$ |
| **124** | **10 May 2011** | **03:40:20** | **49.0079** | **3** | **-25.0** | **-30.0** | **2.07** | **6.4 ± 0.3** | **$1.56\times10^{5}$** |
| 125 | 23 Aug 2011 | 07:39:07 | 149.6889 | 2 | 20.0 | 40.0 | 2.38 | 8.9 ± 0.3 | $1.55\times10^{4}$ |
| 126 | 23 Aug 2011 | 09:58:32 | 149.7822 | 2 | 1.5 | 60.0 | 1.27 | 9.1 ± 0.2 | $1.25\times10^{4}$ |

[a]Also detected by independent observer Dave Clark in Houston, Texas, USA using a 0.2 m Schmidt Cassegrain telescope (Clark 2007).
[b]Also detected by independent observer George Varros in Mt. Airy, Maryland, USA using a 0.2 m Newtonian telescope (Varros 2007, 2008).

**Table 2**
Meteor shower catalog used to constrain lunar impacts. Adapted from Cook (1973) and Jenniskens (1994).

| | $\lambda_{\odot}$ (°) | | | Peak (J2000.0) | | Daily radiant motion | | | |
|---|---|---|---|---|---|---|---|---|---|
| IAU Code | Start | Peak | End | R.A. (°) | Decl. (°) | ΔR.A. (°/day) | ΔDecl. (°/day) | $V_g$ (km/s) | ZHR (#/hr) |
| LYR | 31.4 | 32.4 | 33.4 | 271.4 | 33.6 | 1.1 | 0.0 | 47.6 | 12.8 |
| ETA | 30.7 | 43.1 | 51.7 | 335.6 | -1.9 | 0.9 | 0.4 | 65.5 | 36.7 |
| SDA | 118.7 | 125.7 | 155.7 | 333.1 | -16.6 | 0.8 | 0.2 | 41.4 | 11.4 |
| CAP | 123.7 | 126.7 | 138.7 | 307.0 | -10.0 | 0.9 | 0.3 | 22.8 | 2.2 |
| PER | 120.7 | 139.7 | 150.7 | 46.2 | 57.4 | 1.4 | 0.1 | 59.4 | 84.0 |
| KCG | 136.7 | 145.7 | 193.7 | 286.0 | 59.0 | 0.0 | 0.0 | 24.8 | 2.3 |
| AUR | 155.0 | 158.6 | 162.7 | 84.6 | 42.0 | 1.0 | 0.2 | 66.3 | 9.0 |
| EGE | 201.7 | 206.7 | 214.7 | 104.0 | 27.0 | 0.7 | 0.0 | 69.4 | 2.9 |
| ORI | 189.7 | 208.4 | 225.7 | 94.5 | 15.8 | 1.2 | 0.1 | 66.4 | 25.0 |
| LMI | 209.7 | 211.7 | 211.7 | 162.0 | 37.0 | 1.0 | -0.4 | 61.8 | 1.9 |
| STA | 172.7 | 220.7 | 244.7 | 50.5 | 13.6 | 0.8 | 0.2 | 27.0 | 7.3 |
| NTA | 176.7 | 230.7 | 249.7 | 58.3 | 22.3 | 0.8 | 0.1 | 29.2 | 5.0 |
| LEO | 231.7 | 235.2 | 237.7 | 152.3 | 22.2 | 0.7 | -0.4 | 70.7 | 23.0 |
| PHO | 253.9 | 254.2 | 254.4 | 15.0 | -55.0 | 0.8 | 0.4 | 21.7 | 2.8 |
| MON | 245.7 | 258.7 | 265.7 | 99.8 | 14.0 | 1.0 | -0.1 | 42.4 | 2.0 |
| HYD | 251.7 | 259.7 | 263.7 | 126.6 | 1.6 | 0.7 | -0.2 | 58.4 | 2.5 |
| GEM | 252.7 | 262.4 | 264.9 | 112.3 | 32.5 | 1.0 | -0.1 | 34.4 | 88.0 |
| URS | 265.7 | 270.7 | 272.7 | 217.1 | 75.8 | -0.2 | -0.3 | 33.4 | 11.8 |
| QUA | 281.5 | 283.4 | 284.1 | 230.1 | 48.5 | 0.0 | 0.0 | 41.5 | 120.0 |

**Table 3**
Impactor kinetic energy and size calculations for impact flashes listed in Table 1. Shower association and assumed velocity are shown for each flash alongside kinetic energy and mass estimates using 3 different luminous efficiency assumptions. Impactor diameter is calculated for the velocity dependent luminous efficiency assumption assuming either a shower or average speed (24 km/s). Impact flashes with no convincing shower association are automatically assigned V = 24 km/s. Six flashes which exceeded the dynamic range of the camera as described in Appendix A are indicated in bold. The energies and masses for those are lower limits.

| | | | $\eta = 5\times10^{-4}$ | | $\eta = 1.5\times10^{-3}\exp(-9.3^2/V^2)$ | | | | $\eta = 5\times10^{-3}$ | |
|---|---|---|---|---|---|---|---|---|---|---|
| Flash No. | Shower ID | V (km/s) | Kinetic Energy (J) | Mass (g) | η (V) | Kinetic Energy (J) | Mass (g) | Impactor Diam. (cm) | Kinetic Energy (J) | Mass (g) |
| 1 | | 24.00 | $4.27\times10^{7}$ | 148.3 | $1.29\times10^{-3}$ | $1.65\times10^{7}$ | 57.5 | 4.79 | $4.27\times10^{6}$ | 14.8 |
| 2 | STA | 27.71 | $6.29\times10^{7}$ | 163.8 | $1.34\times10^{-3}$ | $2.35\times10^{7}$ | 61.1 | 4.89 | $6.29\times10^{6}$ | 16.4 |
| 3 | ORI | 66.66 | $9.97\times10^{7}$ | 44.9 | $1.47\times10^{-3}$ | $3.39\times10^{7}$ | 15.3 | 3.08 | $9.97\times10^{6}$ | 4.5 |
| 4 | LEO | 69.95 | $3.12\times10^{7}$ | 12.8 | $1.47\times10^{-3}$ | $1.06\times10^{7}$ | 4.3 | 2.02 | $3.12\times10^{6}$ | 1.3 |
| 5 | LEO | 69.94 | $1.72\times10^{8}$ | 70.2 | $1.47\times10^{-3}$ | $5.82\times10^{7}$ | 23.8 | 3.57 | $1.72\times10^{7}$ | 7.0 |
| 6 | LEO | 69.94 | $1.03\times10^{8}$ | 42.3 | $1.47\times10^{-3}$ | $3.51\times10^{7}$ | 14.3 | 3.01 | $1.03\times10^{7}$ | 4.2 |
| 7 | LEO | 69.94 | $3.86\times10^{7}$ | 15.8 | $1.47\times10^{-3}$ | $1.31\times10^{7}$ | 5.4 | 2.17 | $3.86\times10^{6}$ | 1.6 |
| 8 | NTA | 29.91 | $8.22\times10^{7}$ | 183.7 | $1.36\times10^{-3}$ | $3.02\times10^{7}$ | 67.4 | 5.05 | $8.22\times10^{6}$ | 18.4 |
| 9 | NTA | 29.92 | $7.56\times10^{7}$ | 168.9 | $1.36\times10^{-3}$ | $2.78\times10^{7}$ | 62.0 | 4.91 | $7.56\times10^{6}$ | 16.9 |
| 10 | | 24.00 | $1.13\times10^{8}$ | 393.8 | $1.29\times10^{-3}$ | $4.39\times10^{7}$ | 152.5 | 6.63 | $1.13\times10^{7}$ | 39.4 |
| 11 | NTA | 30.07 | $3.55\times10^{7}$ | 78.6 | $1.36\times10^{-3}$ | $1.30\times10^{7}$ | 28.8 | 3.80 | $3.55\times10^{6}$ | 7.9 |
| 12 | NTA | 30.07 | $3.72\times10^{7}$ | 82.3 | $1.36\times10^{-3}$ | $1.36\times10^{7}$ | 30.2 | 3.86 | $3.72\times10^{6}$ | 8.2 |
| 13 | GEM | 33.44 | $2.14\times10^{7}$ | 38.3 | $1.39\times10^{-3}$ | $7.71\times10^{6}$ | 13.8 | 2.06 | $2.14\times10^{6}$ | 3.8 |
| 14 | GEM | 33.44 | $5.68\times10^{7}$ | 101.7 | $1.39\times10^{-3}$ | $2.05\times10^{7}$ | 36.6 | 2.86 | $5.68\times10^{6}$ | 10.2 |
| 15 | GEM | 33.44 | $3.24\times10^{7}$ | 58.0 | $1.39\times10^{-3}$ | $1.17\times10^{7}$ | 20.9 | 2.37 | $3.24\times10^{6}$ | 5.8 |
| 16 | GEM | 33.44 | $4.60\times10^{7}$ | 82.3 | $1.39\times10^{-3}$ | $1.66\times10^{7}$ | 29.6 | 2.66 | $4.60\times10^{6}$ | 8.2 |
| 17 | GEM | 33.44 | $4.68\times10^{7}$ | 83.8 | $1.39\times10^{-3}$ | $1.69\times10^{7}$ | 30.2 | 2.68 | $4.68\times10^{6}$ | 8.4 |
| 18 | GEM | 33.44 | $1.17\times10^{7}$ | 20.9 | $1.39\times10^{-3}$ | $4.20\times10^{6}$ | 7.5 | 1.68 | $1.17\times10^{6}$ | 2.1 |
| 19 | GEM | 33.44 | $4.43\times10^{7}$ | 79.3 | $1.39\times10^{-3}$ | $1.60\times10^{7}$ | 28.6 | 2.63 | $4.43\times10^{6}$ | 7.9 |
| 20 | GEM | 33.44 | $1.21\times10^{8}$ | 216.3 | $1.39\times10^{-3}$ | $4.36\times10^{7}$ | 77.9 | 3.67 | $1.21\times10^{7}$ | 21.6 |
| 21 | | 24.00 | $7.22\times10^{7}$ | 250.8 | $1.29\times10^{-3}$ | $2.80\times10^{7}$ | 97.1 | 5.70 | $7.22\times10^{6}$ | 25.1 |
| 22 | | 24.00 | $2.65\times10^{7}$ | 91.9 | $1.29\times10^{-3}$ | $1.03\times10^{7}$ | 35.6 | 4.08 | $2.65\times10^{6}$ | 9.2 |
| 23 | | 24.00 | $1.99\times10^{7}$ | 69.1 | $1.29\times10^{-3}$ | $7.70\times10^{6}$ | 26.8 | 3.71 | $1.99\times10^{6}$ | 6.9 |
| 24 | | 24.00 | $2.04\times10^{7}$ | 71.0 | $1.29\times10^{-3}$ | $7.92\times10^{6}$ | 27.5 | 3.74 | $2.04\times10^{6}$ | 7.1 |
| 25 | LYR | 47.57 | $9.88\times10^{7}$ | 87.3 | $1.44\times10^{-3}$ | $3.42\times10^{7}$ | 30.2 | 3.87 | $9.88\times10^{6}$ | 8.7 |
| 26 | LYR | 47.58 | $2.22\times10^{8}$ | 196.3 | $1.44\times10^{-3}$ | $7.69\times10^{7}$ | 68.0 | 5.06 | $2.22\times10^{7}$ | 19.6 |
| 27 | LYR | 47.58 | $4.12\times10^{7}$ | 36.4 | $1.44\times10^{-3}$ | $1.43\times10^{7}$ | 12.6 | 2.89 | $4.12\times10^{6}$ | 3.6 |
| 28 | LYR | 47.59 | $5.90\times10^{7}$ | 52.1 | $1.44\times10^{-3}$ | $2.04\times10^{7}$ | 18.0 | 3.25 | $5.90\times10^{6}$ | 5.2 |
| 29 | LYR | 47.67 | $5.28\times10^{7}$ | 46.5 | $1.44\times10^{-3}$ | $1.83\times10^{7}$ | 16.1 | 3.13 | $5.28\times10^{6}$ | 4.6 |
| 30 | ETA | 64.81 | $3.21\times10^{7}$ | 15.3 | $1.47\times10^{-3}$ | $1.09\times10^{7}$ | 5.2 | 2.15 | $3.21\times10^{6}$ | 1.5 |
| 31 | | 24.00 | $3.93\times10^{7}$ | 136.5 | $1.29\times10^{-3}$ | $1.52\times10^{7}$ | 52.9 | 4.66 | $3.93\times10^{6}$ | 13.7 |
| 32 | LYR | 47.68 | $6.65\times10^{7}$ | 58.5 | $1.44\times10^{-3}$ | $2.30\times10^{7}$ | 20.2 | 3.38 | $6.65\times10^{6}$ | 5.8 |
| 33 | LYR | 47.68 | $4.12\times10^{7}$ | 36.2 | $1.44\times10^{-3}$ | $1.43\times10^{7}$ | 12.5 | 2.88 | $4.12\times10^{6}$ | 3.6 |
| 34 | LYR | 47.68 | $2.04\times10^{7}$ | 18.0 | $1.44\times10^{-3}$ | $7.08\times10^{6}$ | 6.2 | 2.28 | $2.04\times10^{6}$ | 1.8 |
| 35 | LYR | 47.68 | $1.69\times10^{8}$ | 148.3 | $1.44\times10^{-3}$ | $5.84\times10^{7}$ | 51.3 | 4.61 | $1.69\times10^{7}$ | 14.8 |
| 36 | LYR | 47.69 | $9.26\times10^{7}$ | 81.4 | $1.44\times10^{-3}$ | $3.21\times10^{7}$ | 28.2 | 3.78 | $9.26\times10^{6}$ | 8.1 |
| 37 | | 24.00 | $7.99\times10^{7}$ | 277.5 | $1.29\times10^{-3}$ | $3.10\times10^{7}$ | 107.5 | 5.90 | $7.99\times10^{6}$ | 27.7 |
| 38 | | 24.00 | $2.67\times10^{7}$ | 92.7 | $1.29\times10^{-3}$ | $1.03\times10^{7}$ | 35.9 | 4.09 | $2.67\times10^{6}$ | 9.3 |
| 39 | SDA | 40.45 | $1.29\times10^{8}$ | 157.7 | $1.42\times10^{-3}$ | $4.53\times10^{7}$ | 55.4 | 4.73 | $1.29\times10^{7}$ | 15.8 |
| 40 | STA | 26.04 | $4.04\times10^{7}$ | 119.2 | $1.32\times10^{-3}$ | $1.53\times10^{7}$ | 45.2 | 4.42 | $4.04\times10^{6}$ | 11.9 |
| 41 | NTA | 30.07 | $2.88\times10^{7}$ | 63.6 | $1.36\times10^{-3}$ | $1.05\times10^{7}$ | 23.3 | 3.54 | $2.88\times10^{6}$ | 6.4 |
| **42** | **NTA** | **30.07** | **$1.62\times10^{8}$** | **359.3** | **$1.36\times10^{-3}$** | **$5.96\times10^{7}$** | **131.8** | **6.31** | **$1.62\times10^{7}$** | **35.9** |
| 43 | QUA | 41.30 | $2.39\times10^{7}$ | 28.0 | $1.43\times10^{-3}$ | $8.39\times10^{6}$ | 9.8 | 1.84 | $2.39\times10^{6}$ | 2.8 |
| 44 | QUA | 41.23 | $5.84\times10^{7}$ | 68.7 | $1.43\times10^{-3}$ | $2.05\times10^{7}$ | 24.1 | 2.49 | $5.84\times10^{6}$ | 6.9 |
| **45** | **QUA** | **41.23** | **$1.51\times10^{8}$** | **177.5** | **$1.43\times10^{-3}$** | **$5.29\times10^{7}$** | **62.3** | **3.41** | **$1.51\times10^{7}$** | **17.8** |
| 46 | QUA | 41.23 | $5.33\times10^{7}$ | 62.7 | $1.43\times10^{-3}$ | $1.87\times10^{7}$ | 22.0 | 2.41 | $5.33\times10^{6}$ | 6.3 |
| 47 | | 24.00 | $1.50\times10^{7}$ | 51.9 | $1.29\times10^{-3}$ | $5.79\times10^{6}$ | 20.1 | 3.37 | $1.50\times10^{6}$ | 5.2 |
| 48 | | 24.00 | $5.79\times10^{7}$ | 201.0 | $1.29\times10^{-3}$ | $2.24\times10^{7}$ | 77.9 | 5.30 | $5.79\times10^{6}$ | 20.1 |
| 49 | | 24.00 | $4.04\times10^{7}$ | 140.4 | $1.29\times10^{-3}$ | $1.57\times10^{7}$ | 54.4 | 4.70 | $4.04\times10^{6}$ | 14.0 |
| **50** | | **24.00** | **$2.06\times10^{8}$** | **716.6** | **$1.29\times10^{-3}$** | **$7.99\times10^{7}$** | **277.5** | **8.09** | **$2.06\times10^{7}$** | **71.7** |
| 51 | | 24.00 | $2.39\times10^{7}$ | 83.0 | $1.29\times10^{-3}$ | $9.26\times10^{6}$ | 32.2 | 3.95 | $2.39\times10^{6}$ | 8.3 |
| 52 | | 24.00 | $2.10\times10^{7}$ | 73.0 | $1.29\times10^{-3}$ | $8.14\times10^{6}$ | 28.3 | 3.78 | $2.10\times10^{6}$ | 7.3 |
| 53 | | 24.00 | $7.36\times10^{6}$ | 25.5 | $1.29\times10^{-3}$ | $2.85\times10^{6}$ | 9.9 | 2.66 | $7.36\times10^{5}$ | 2.6 |
| **54** | | **24.00** | **$9.18\times10^{7}$** | **318.6** | **$1.29\times10^{-3}$** | **$3.55\times10^{7}$** | **123.4** | **6.18** | **$9.18\times10^{6}$** | **31.9** |
| 55 | | 24.00 | $4.04\times10^{8}$ | 1403.6 | $1.29\times10^{-3}$ | $1.57\times10^{8}$ | 543.7 | 10.13 | $4.04\times10^{7}$ | 140.4 |
| 56 | | 24.00 | $4.68\times10^{7}$ | 162.6 | $1.29\times10^{-3}$ | $1.81\times10^{7}$ | 63.0 | 4.94 | $4.68\times10^{6}$ | 16.3 |
| 57 | | 24.00 | $2.50\times10^{8}$ | 869.5 | $1.29\times10^{-3}$ | $9.70\times10^{7}$ | 336.8 | 8.63 | $2.50\times10^{7}$ | 86.9 |
| 58 | | 24.00 | $7.22\times10^{7}$ | 250.8 | $1.29\times10^{-3}$ | $2.80\times10^{7}$ | 97.1 | 5.70 | $7.22\times10^{6}$ | 25.1 |
| 59 | | 24.00 | $3.69\times10^{7}$ | 128.0 | $1.29\times10^{-3}$ | $1.43\times10^{7}$ | 49.6 | 4.56 | $3.69\times10^{6}$ | 12.8 |
| 60 | SDA | 40.36 | $9.61\times10^{6}$ | 11.8 | $1.42\times10^{-3}$ | $3.38\times10^{6}$ | 4.1 | 1.99 | $9.61\times10^{5}$ | 1.2 |
| 61 | SDA | 40.46 | $3.21\times10^{7}$ | 39.2 | $1.42\times10^{-3}$ | $1.13\times10^{7}$ | 13.8 | 2.97 | $3.21\times10^{6}$ | 3.9 |
| 62 | SDA | 40.36 | $4.43\times10^{7}$ | 54.4 | $1.42\times10^{-3}$ | $1.56\times10^{7}$ | 19.1 | 3.32 | $4.43\times10^{6}$ | 5.4 |
| 63 | SDA | 40.36 | $7.92\times10^{7}$ | 97.2 | $1.42\times10^{-3}$ | $2.78\times10^{7}$ | 34.2 | 4.03 | $7.92\times10^{6}$ | 9.7 |
| 64 | | 24.00 | $2.93\times10^{7}$ | 101.7 | $1.29\times10^{-3}$ | $1.13\times10^{7}$ | 39.4 | 4.22 | $2.93\times10^{6}$ | 10.2 |
| 65 | | 24.00 | $3.15\times10^{8}$ | 1094.6 | $1.29\times10^{-3}$ | $1.22\times10^{8}$ | 424.0 | 9.32 | $3.15\times10^{7}$ | 109.5 |
| 66 | STA | 25.98 | $6.90\times10^{7}$ | 204.4 | $1.32\times10^{-3}$ | $2.61\times10^{7}$ | 77.4 | 5.29 | $6.90\times10^{6}$ | 20.4 |
| 67 | ORI | 65.87 | $4.51\times10^{7}$ | 20.8 | $1.47\times10^{-3}$ | $1.54\times10^{7}$ | 7.1 | 2.38 | $4.51\times10^{6}$ | 2.1 |
| 68 | ORI | 65.85 | $2.80\times10^{7}$ | 12.9 | $1.47\times10^{-3}$ | $9.51\times10^{6}$ | 4.4 | 2.03 | $2.80\times10^{6}$ | 1.3 |
| 69 | ORI | 65.85 | $2.88\times10^{7}$ | 13.3 | $1.47\times10^{-3}$ | $9.78\times10^{6}$ | 4.5 | 2.05 | $2.88\times10^{6}$ | 1.3 |
| 70 | STA | 27.67 | $3.15\times10^{7}$ | 82.3 | $1.34\times10^{-3}$ | $1.18\times10^{7}$ | 30.7 | 3.89 | $3.15\times10^{6}$ | 8.2 |
| 71 | STA | 27.67 | $8.76\times10^{7}$ | 228.9 | $1.34\times10^{-3}$ | $3.27\times10^{7}$ | 85.4 | 5.46 | $8.76\times10^{6}$ | 22.9 |
| 72 | STA | 27.80 | $2.90\times10^{7}$ | 75.1 | $1.34\times10^{-3}$ | $1.08\times10^{7}$ | 28.0 | 3.77 | $2.90\times10^{6}$ | 7.5 |
| 73 | STA | 27.80 | $4.39\times10^{7}$ | 113.7 | $1.34\times10^{-3}$ | $1.64\times10^{7}$ | 42.4 | 4.33 | $4.39\times10^{6}$ | 11.4 |
| 74 | STA | 27.80 | $4.31\times10^{7}$ | 111.6 | $1.34\times10^{-3}$ | $1.61\times10^{7}$ | 41.6 | 4.30 | $4.31\times10^{6}$ | 11.2 |
| 75 | STA | 27.80 | $1.12\times10^{8}$ | 290.8 | $1.34\times10^{-3}$ | $4.19\times10^{7}$ | 108.4 | 5.92 | $1.12\times10^{7}$ | 29.1 |
| 76 | STA | 27.89 | $1.06\times10^{8}$ | 273.4 | $1.34\times10^{-3}$ | $3.96\times10^{7}$ | 101.8 | 5.79 | $1.06\times10^{7}$ | 27.3 |
| 77 | STA | 27.90 | $2.75\times10^{7}$ | 70.5 | $1.34\times10^{-3}$ | $1.02\times10^{7}$ | 26.3 | 3.69 | $2.75\times10^{6}$ | 7.1 |
| 78 | STA | 27.90 | $1.50\times10^{8}$ | 384.1 | $1.34\times10^{-3}$ | $5.57\times10^{7}$ | 143.1 | 6.49 | $1.50\times10^{7}$ | 38.4 |
| 79 | STA | 27.90 | $1.88\times10^{7}$ | 48.4 | $1.34\times10^{-3}$ | $7.01\times10^{6}$ | 18.0 | 3.25 | $1.88\times10^{6}$ | 4.8 |
| 80 | STA | 27.90 | $7.16\times10^{7}$ | 183.8 | $1.34\times10^{-3}$ | $2.67\times10^{7}$ | 68.5 | 5.08 | $7.16\times10^{6}$ | 18.4 |
| 81 | NTA | 28.18 | $3.59\times10^{7}$ | 90.3 | $1.35\times10^{-3}$ | $1.33\times10^{7}$ | 33.6 | 4.00 | $3.59\times10^{6}$ | 9.0 |
| 82 | NTA | 28.26 | $8.22\times10^{7}$ | 205.7 | $1.35\times10^{-3}$ | $3.05\times10^{7}$ | 76.4 | 5.27 | $8.22\times10^{6}$ | 20.6 |
| 83 | NTA | 28.36 | $3.49\times10^{7}$ | 86.7 | $1.35\times10^{-3}$ | $1.29\times10^{7}$ | 32.2 | 3.95 | $3.49\times10^{6}$ | 8.7 |
| 84 | NTA | 28.37 | $2.50\times10^{7}$ | 62.2 | $1.35\times10^{-3}$ | $9.29\times10^{6}$ | 23.1 | 3.53 | $2.50\times10^{6}$ | 6.2 |
| 85 | MON | 42.98 | $1.02\times10^{7}$ | 11.0 | $1.43\times10^{-3}$ | $3.55\times10^{6}$ | 3.8 | 1.94 | $1.02\times10^{6}$ | 1.1 |
| 86 | MON | 42.99 | $7.36\times10^{7}$ | 79.6 | $1.43\times10^{-3}$ | $2.57\times10^{7}$ | 27.8 | 3.76 | $7.36\times10^{6}$ | 8.0 |
| 87 | | 24.00 | $1.93\times10^{7}$ | 67.2 | $1.29\times10^{-3}$ | $7.49\times10^{6}$ | 26.0 | 3.68 | $1.93\times10^{6}$ | 6.7 |
| 88 | | 24.00 | $2.98\times10^{7}$ | 103.6 | $1.29\times10^{-3}$ | $1.16\times10^{7}$ | 40.1 | 4.25 | $2.98\times10^{6}$ | 10.4 |
| 89 | | 24.00 | $3.24\times10^{7}$ | 112.5 | $1.29\times10^{-3}$ | $1.26\times10^{7}$ | 43.6 | 4.37 | $3.24\times10^{6}$ | 11.3 |
| 90 | | 24.00 | $2.18\times10^{7}$ | 75.7 | $1.29\times10^{-3}$ | $8.45\times10^{6}$ | 29.3 | 3.83 | $2.18\times10^{6}$ | 7.6 |
| 91 | | 24.00 | $3.59\times10^{7}$ | 124.5 | $1.29\times10^{-3}$ | $1.39\times10^{7}$ | 48.2 | 4.52 | $3.59\times10^{6}$ | 12.5 |
| 92 | | 24.00 | $6.96\times10^{7}$ | 241.7 | $1.29\times10^{-3}$ | $2.70\times10^{7}$ | 93.6 | 5.63 | $6.96\times10^{6}$ | 24.2 |
| 93 | | 24.00 | $5.00\times10^{7}$ | 173.5 | $1.29\times10^{-3}$ | $1.94\times10^{7}$ | 67.2 | 5.04 | $5.00\times10^{6}$ | 17.3 |
| 94 | | 24.00 | $2.12\times10^{8}$ | 736.6 | $1.29\times10^{-3}$ | $8.22\times10^{7}$ | 285.3 | 8.17 | $2.12\times10^{7}$ | 73.7 |
| 95 | | 24.00 | $1.13\times10^{8}$ | 393.8 | $1.29\times10^{-3}$ | $4.39\times10^{7}$ | 152.5 | 6.63 | $1.13\times10^{7}$ | 39.4 |
| 96 | | 24.00 | $9.79\times10^{7}$ | 339.8 | $1.29\times10^{-3}$ | $3.79\times10^{7}$ | 131.6 | 6.31 | $9.79\times10^{6}$ | 34.0 |
| 97 | ORI | 66.58 | $3.59\times10^{7}$ | 16.2 | $1.47\times10^{-3}$ | $1.22\times10^{7}$ | 5.5 | 2.19 | $3.59\times10^{6}$ | 1.6 |
| 98 | ORI | 66.58 | $7.77\times10^{6}$ | 3.5 | $1.47\times10^{-3}$ | $2.64\times10^{6}$ | 1.2 | 1.32 | $7.77\times10^{5}$ | 0.4 |
| 99 | ORI | 66.59 | $3.18\times10^{7}$ | 14.4 | $1.47\times10^{-3}$ | $1.08\times10^{7}$ | 4.9 | 2.10 | $3.18\times10^{6}$ | 1.4 |
| 100 | ORI | 66.59 | $1.78\times10^{7}$ | 8.0 | $1.47\times10^{-3}$ | $6.05\times10^{6}$ | 2.7 | 1.73 | $1.78\times10^{6}$ | 0.8 |
| 101 | ORI | 66.59 | $3.86\times10^{7}$ | 17.4 | $1.47\times10^{-3}$ | $1.31\times10^{7}$ | 5.9 | 2.24 | $3.86\times10^{6}$ | 1.7 |
| 102 | ORI | 66.59 | $6.71\times10^{7}$ | 30.3 | $1.47\times10^{-3}$ | $2.28\times10^{7}$ | 10.3 | 2.70 | $6.71\times10^{6}$ | 3.0 |
| 103 | ORI | 66.60 | $3.55\times10^{7}$ | 16.0 | $1.47\times10^{-3}$ | $1.21\times10^{7}$ | 5.4 | 2.18 | $3.55\times10^{6}$ | 1.6 |
| 104 | ORI | 66.60 | $4.31\times10^{7}$ | 19.4 | $1.47\times10^{-3}$ | $1.47\times10^{7}$ | 6.6 | 2.33 | $4.31\times10^{6}$ | 1.9 |
| 105 | ORI | 66.60 | $6.96\times10^{7}$ | 31.4 | $1.47\times10^{-3}$ | $2.37\times10^{7}$ | 10.7 | 2.73 | $6.96\times10^{6}$ | 3.1 |
| 106 | NTA | 28.24 | $4.68\times10^{7}$ | 117.5 | $1.35\times10^{-3}$ | $1.74\times10^{7}$ | 43.6 | 4.37 | $4.68\times10^{6}$ | 11.7 |
| 107 | NTA | 28.37 | $7.56\times10^{7}$ | 187.9 | $1.35\times10^{-3}$ | $2.81\times10^{7}$ | 69.7 | 5.11 | $7.56\times10^{6}$ | 18.8 |
| 108 | GEM | 33.40 | $4.47\times10^{7}$ | 80.2 | $1.39\times10^{-3}$ | $1.61\times10^{7}$ | 28.9 | 2.64 | $4.47\times10^{6}$ | 8.0 |
| 109 | URS | 33.54 | $9.52\times10^{7}$ | 169.3 | $1.39\times10^{-3}$ | $3.43\times10^{7}$ | 60.9 | 4.88 | $9.52\times10^{6}$ | 16.9 |
| 110 | | 24.00 | $4.16\times10^{7}$ | 144.3 | $1.29\times10^{-3}$ | $1.61\times10^{7}$ | 55.9 | 4.74 | $4.16\times10^{6}$ | 14.4 |
| 111 | | 24.00 | $3.12\times10^{7}$ | 108.5 | $1.29\times10^{-3}$ | $1.21\times10^{7}$ | 42.0 | 4.31 | $3.12\times10^{6}$ | 10.8 |
| 112 | | 24.00 | $4.82\times10^{7}$ | 167.2 | $1.29\times10^{-3}$ | $1.87\times10^{7}$ | 64.8 | 4.98 | $4.82\times10^{6}$ | 16.7 |
| 113 | | 24.00 | $2.98\times10^{7}$ | 103.6 | $1.29\times10^{-3}$ | $1.16\times10^{7}$ | 40.1 | 4.25 | $2.98\times10^{6}$ | 10.4 |
| 114 | | 24.00 | $2.80\times10^{7}$ | 97.1 | $1.29\times10^{-3}$ | $1.08\times10^{7}$ | 37.6 | 4.16 | $2.80\times10^{6}$ | 9.7 |
| 115 | | 24.00 | $1.41\times10^{7}$ | 49.1 | $1.29\times10^{-3}$ | $5.48\times10^{6}$ | 19.0 | 3.31 | $1.41\times10^{6}$ | 4.9 |
| 116 | | 24.00 | $1.04\times10^{8}$ | 362.5 | $1.29\times10^{-3}$ | $4.04\times10^{7}$ | 140.4 | 6.45 | $1.04\times10^{7}$ | 36.2 |
| **117** | | **24.00** | $\mathbf{1.02\times10^{9}}$ | **3525.7** | $\mathbf{1.29\times10^{-3}}$ | $\mathbf{3.93\times10^{8}}$ | **1365.7** | **13.77** | $\mathbf{1.02\times10^{8}}$ | **352.6** |
| 118 | | 24.00 | $5.00\times10^{7}$ | 173.5 | $1.29\times10^{-3}$ | $1.94\times10^{7}$ | 67.2 | 5.04 | $5.00\times10^{6}$ | 17.3 |
| 119 | STA | 26.04 | $5.84\times10^{7}$ | 172.3 | $1.32\times10^{-3}$ | $2.21\times10^{7}$ | 65.3 | 5.00 | $5.84\times10^{6}$ | 17.2 |
| 120 | | 24.00 | $4.77\times10^{7}$ | 165.7 | $1.29\times10^{-3}$ | $1.85\times10^{7}$ | 64.2 | 4.97 | $4.77\times10^{6}$ | 16.6 |
| 121 | | 24.00 | $2.96\times10^{7}$ | 102.6 | $1.29\times10^{-3}$ | $1.14\times10^{7}$ | 39.8 | 4.23 | $2.96\times10^{6}$ | 10.3 |
| 122 | | 24.00 | $3.04\times10^{7}$ | 105.5 | $1.29\times10^{-3}$ | $1.18\times10^{7}$ | 40.9 | 4.27 | $3.04\times10^{6}$ | 10.5 |
| 123 | | 24.00 | $1.23\times10^{8}$ | 427.8 | $1.29\times10^{-3}$ | $4.77\times10^{7}$ | 165.7 | 6.81 | $1.23\times10^{7}$ | 42.8 |
| **124** | **ETA** | **64.98** | $\mathbf{3.12\times10^{8}}$ | **147.9** | $\mathbf{1.47\times10^{-3}}$ | $\mathbf{1.06\times10^{8}}$ | **50.3** | **4.58** | $\mathbf{3.12\times10^{7}}$ | **14.8** |
| 125 | SDA | 40.43 | $3.09\times10^{7}$ | 37.9 | $1.42\times10^{-3}$ | $1.09\times10^{7}$ | 13.3 | 2.94 | $3.09\times10^{6}$ | 3.8 |
| 126 | SDA | 40.43 | $2.50\times10^{7}$ | 30.6 | $1.42\times10^{-3}$ | $8.80\times10^{6}$ | 10.8 | 2.74 | $2.50\times10^{6}$ | 3.1 |